\documentclass[10pt,twocolumn]{article}
\usepackage{amssymb}
\usepackage{graphicx}
\usepackage{epsfig}
\usepackage{rotating} 
\usepackage{longtable}
\usepackage{afterpage}
\def\Sigbar{\ensuremath{\kern 0.2em\overline{\kern -0.2em \Sigma}}{}}
\def\Lbar {\ensuremath{\kern 0.2em\overline{\kern -0.2em\Lambda\kern 0.05em}\kern-0.05em}{}}

\begin{document}

\title{\bf\boldmath Study of $e^+e^-$ annihilation at low energies}
\author{V.P.~Druzhinin\\ \itshape Budker Institute of Nuclear Physics,
Novosibirsk 630090, Russia}
\maketitle
\begin{abstract}
The recent results of the CMD-2, SND, KLOE, and BABAR experiments on $e^+e^-$ 
annihilation into hadrons at low energies are reviewed.
\end{abstract}

\section{Introduction}\label{intro}
The $e^+e^-$ annihilation at low energies, below 3 GeV, is the main
source of information on the properties  of the light vector mesons
($\rho$, $\omega$, $\phi$) and their excited states. 
Other important application
of the $e^+e^-$ annihilation is calculation of the hadronic contribution into
the anomalous magnetic moment of muon $a_\mu^{\rm had}$ and the running 
electromagnetic constant. The difference between the measured value 
of $a_\mu$~\cite{g-2-ex} and recent Standard model 
calculations~\cite{g-2-th1,g-2-th2,g-2-th3} varies from 3.2 to 3.4 sigma.
Currently the hadronic leading-order contribution into $a_\mu$ is dominant
source of uncertainty of the Standard model $a_\mu$ prediction. It is 
evaluated using experimental data on $e^+e^-$ annihilation via
dispersion integral
$$a_\mu^{\rm had}=\int_{4m_\pi^2}^{\infty} {\rm d}s K(s) R(s),$$
where $R(s)$ is the ratio of the total cross section of the $e^+e^-$
annihilation into hadrons to the $e^+e^-\to \mu^+\mu^-$ cross section, and
$K(s)$ is the kernel function decreasing monotonically as approximately
$1/s^2$. Due to such $s$-dependence of $K(s)$ the main contribution into
$a_\mu^{\rm had}$ comes from the low energy region.
Table~\ref{tab1} shows the contributions of the different $e^+e^-$ 
center-of-mass (c.m.) energy regions and the different processes into 
the $a_\mu^{\rm had}$~\cite{g-2-th1}.
\begin{table*}
\caption{\label{tab1} The contributions of different final states to $a_\mu^{\rm had}$,
given in units of $10^{-10}$.}
\begin{center}
\begin{tabular}{lc}
\hline
Modes               &  $a_\mu^{\rm had}$  \\
$\pi^+\pi^-$        &  $504.6 \pm 3.1 \pm 1.0_{\rm rad}$ \\
$4\pi$              &  $29.9  \pm 1.4 \pm 0.2_{\rm rad}$ \\
$\omega$            &  $38.0  \pm 1.0 \pm 0.3_{\rm rad}$ \\
$\phi$              &  $35.7  \pm 0.8 \pm 0.2_{\rm rad}$ \\
other, $E < 1.8$ GeV&  $24.3  \pm 1.3 \pm 0.2_{\rm rad}$ \\
$E > 1.8$ GeV       &  $58.4  \pm 0.5 \pm 0.7_{\rm QCD}$ \\ \hline
sum                 &  $690.8 \pm 3.9 \pm 1.9_{\rm rad} \pm 0.7_{\rm QCD}$ \\
\hline
\end{tabular}
\end{center}
\end{table*}
It is seen that the dominant contribution both into the $a_\mu^{\rm had}$ value
and into its uncertainty comes from the energy region below 1.8 GeV and
from the process $e^+e^-\to \pi^+\pi^-$.

\begin{figure*}
\includegraphics[width=.9\linewidth]{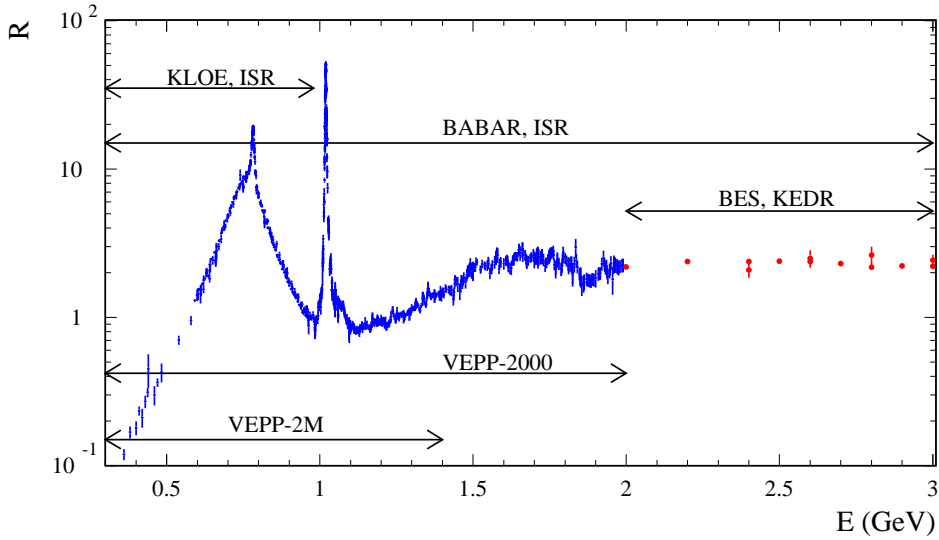}
\caption{The $R$ dependence on $e^+e^-$ center of mass energy.
The arrows indicate the energy regions covered by the present and future
experiments contributing into the $R$ measurement at low energies.
\label{rvse}}
\end{figure*}
The $R$ energy dependence~\cite{pdg} for energies below 3 GeV is shown in Fig.\ref{rvse}.
The arrows indicate the energy regions covered by the present and future
experiments contributing into the $R$ measurement at low energies. In the resonant
energy region, below 2 GeV, the total cross section is obtained by summing
the exclusive cross sections for different hadronic final modes. For higher
energies the inclusive approach is used. For $E < 1.4$ GeV, the most full and
accurate data were obtained in CMD-2 and SND experiments at VEPP-2M. The 
VEPP-2M $e^+e^-$ collider finished to work in 2000. Now this machine is substituted 
by the new collider VEPP-2000 with maximal energy of 2 GeV. The experiments at
VEPP-2000 with SND and CMD-3 detectors will start in 2008. The most accurate
measurement of the total hadronic cross section in the 2--3 GeV range was performed
by BES detector~\cite{BES}. New measurement in this energy range is planned at
VEPP-4 $e^+e^-$ collider with KEDR detector.

\begin{figure}
\begin{center}
\includegraphics[width=.4\textwidth]{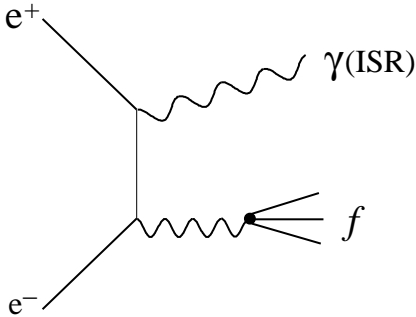}
\caption{The diagram of the $e^+e^-\to f\gamma$ process, where $f$ is
hadronic system.
\label{diag3}}
\end{center}
\end{figure}

The relatively new initial state radiation (ISR) or radiative return technique
is used for measurements of exclusive hadronic cross sections at 
high luminosity $e^+e^-$ factories. The Born cross section for the ISR process
$e^+e^-\to f + \gamma$ (Fig.\ref{diag3}), where $f$ is hadronic system,
integrated over the hadron momenta, is given by
$$
\frac{{\rm d}\sigma_{e^+e^-\to f \gamma}(m)}
{{\rm d}m\,{\rm d}\cos{\theta_\gamma}} = 
\frac{2m}{s}\, W(s,x,\theta_\gamma)\,\sigma_{f}(m),
$$
where $\sqrt{s}$ is the $e^+e^-$ c.m. energy,
$m$ is the invariant mass of the hadronic system,
$\sigma_{f}(m)$ is the cross section for $e^+e^-\to f$ reaction,
$x\equiv{E_{\gamma}}/\sqrt{s}=1-{m^2}/{s}$, $E_{\gamma}$ and
$\theta_\gamma$
are the ISR photon energy and polar angle, respectively,  
in the $e^+e^-$ c.m. frame.
The function $W(s,x,\theta_{\gamma})$~\cite{BM}
describes the probability of the ISR photon emission.
The ISR photon is predominantly emitted along beam axis. 
Two approaches are used for ISR measurements, with and without detection of
the photon. The approach with untagged photon, in principle, provides higher 
detection efficiency for ISR events,
but this efficiency strongly depends on the invariant mass of the hadronic
system. The finite detector acceptance does not allow to 
use this approach for measurements in the near-threshold mass region.
The approach with detected photon ($30^\circ < \theta_\gamma <150^\circ$) 
utilizes only about 10\% of ISR events but provides the weak dependence of 
the detection efficiency on the mass of the hadronic system and its internal
substructure. The untagged approach is used by KLOE detector at $\phi$-factory
to measure the $e^+e^-\to \pi^+\pi^-$ cross section. The approach with detected
photon is used by BABAR at $B$-factory ($\sqrt(s)=10.6$ GeV).
\section{VEPP-2M results}
\begin{figure*}
\includegraphics[width=.9\linewidth]{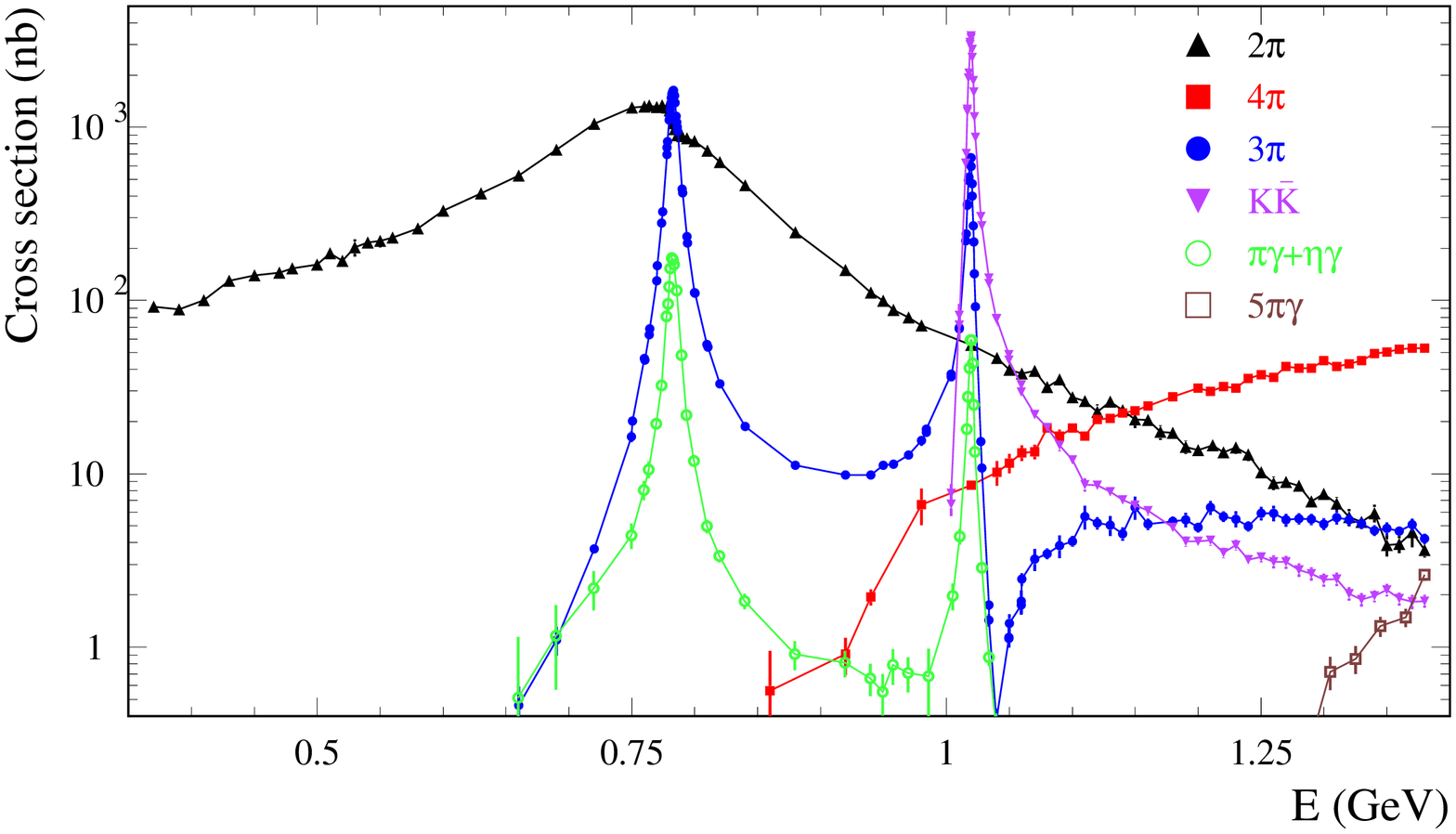}
\caption{The hadronic cross sections measured by CMD-2 and SND detectors
at VEPP-2M collider.
\label{vepp2m}}
\end{figure*}

The CMD-2 and SND detectors took data at the VEPP-2M $e^+e^-$ collider
during 1995--2000.
The total integrated luminosity collected by the two detectors in the energy
range 0.36-1.4 GeV is about 60 pb$^{-1}$. All major hadronic cross sections
were measured. The latest published results are measurements of 
the $e^+e^-\to K^+K^-$~\cite{kkc_snd} and $e^+e^-\to \eta\gamma$~\cite{etag_snd}
cross sections at SND. Analyses of some important modes are in progress. In
particular, new results on $e^+e^-\to 3\pi,\, \pi^+\pi^-2\pi^0,\, K^+K^-$ are
expected from CMD-2, and $e^+e^-\to 2\pi$ for $E > 1$ GeV and $e^+e^-\to
\pi^0\gamma$ from SND. 
The relative contributions of different processes measured at 
VEPP-2M and achieved statistical accuracy of the cross-section measurements
are demonstrated in Fig.~\ref{vepp2m}. 

\begin{figure*}
\includegraphics[width=.9\linewidth]{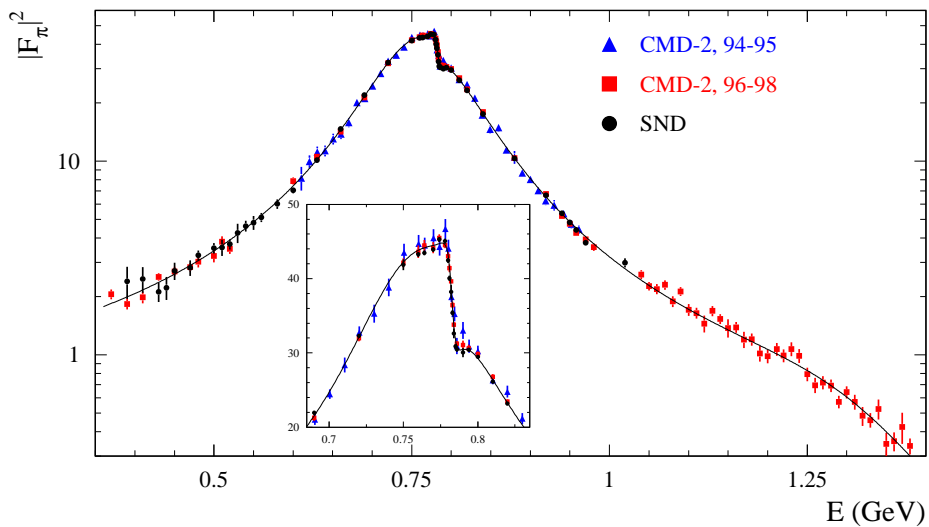}
\caption{The pion form factor measured by CMD-2 and SND detectors.
The curve is a fit result. The fit takes into account contributions 
of $\rho$, $\omega$, and $\rho^\prime$ resonances.
\label{2pi_vepp2m}}
\end{figure*}

The most important process for calculation of the $a_\mu^{\rm had}$
is $e^+e^-\to 2\pi$.
At CMD-2 this process was measured in the energy range from 0.37 GeV up to
1.38 GeV~\cite{2pi_cmd_1,2pi_cmd_2,2pi_cmd_3,2pi_cmd_4}. The systematic
uncertainty of the CMD-2 measurement is 0.6--0.8\% for $E < 1$ GeV. For
energies above 1 GeV it varies from 1.2\% to 4.2\%. The SND measured 
the $e^+e^-\to 2\pi$ cross section in the energy range 0.4--1.0 GeV with the
systematic uncertainty of 1.3\%~\cite{2pi_snd_1}. There is also SND measurement
in the $\phi$-meson vicinity~\cite{2pi_snd_1}. The CMD-2 and SND results are
compared in Fig.~\ref{2pi_vepp2m}.
The data of two detectors are in good agreement.
\section{KLOE results}
\begin{figure*}
\includegraphics[width=.45\linewidth]{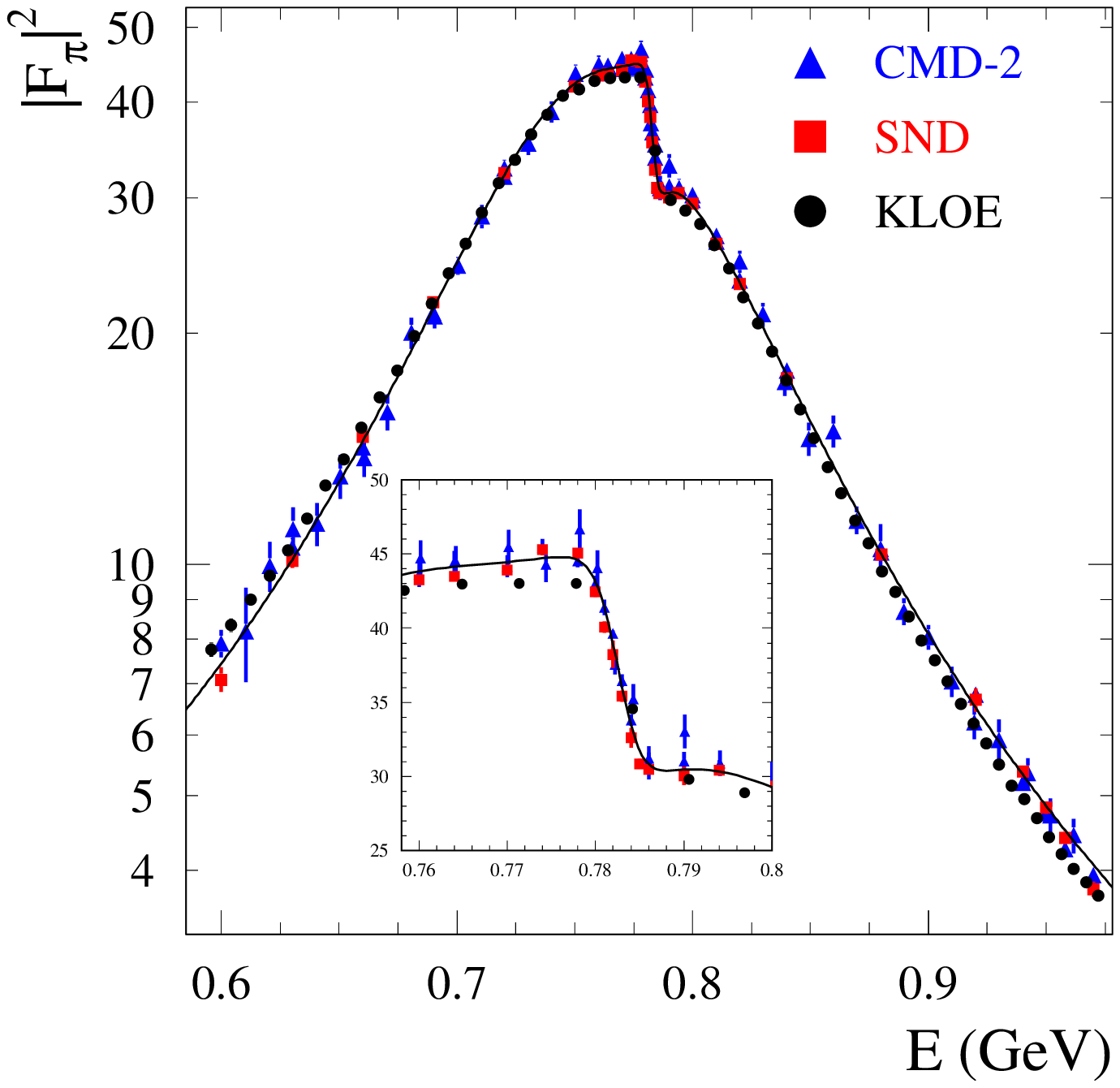}
\hfill
\includegraphics[width=.45\linewidth]{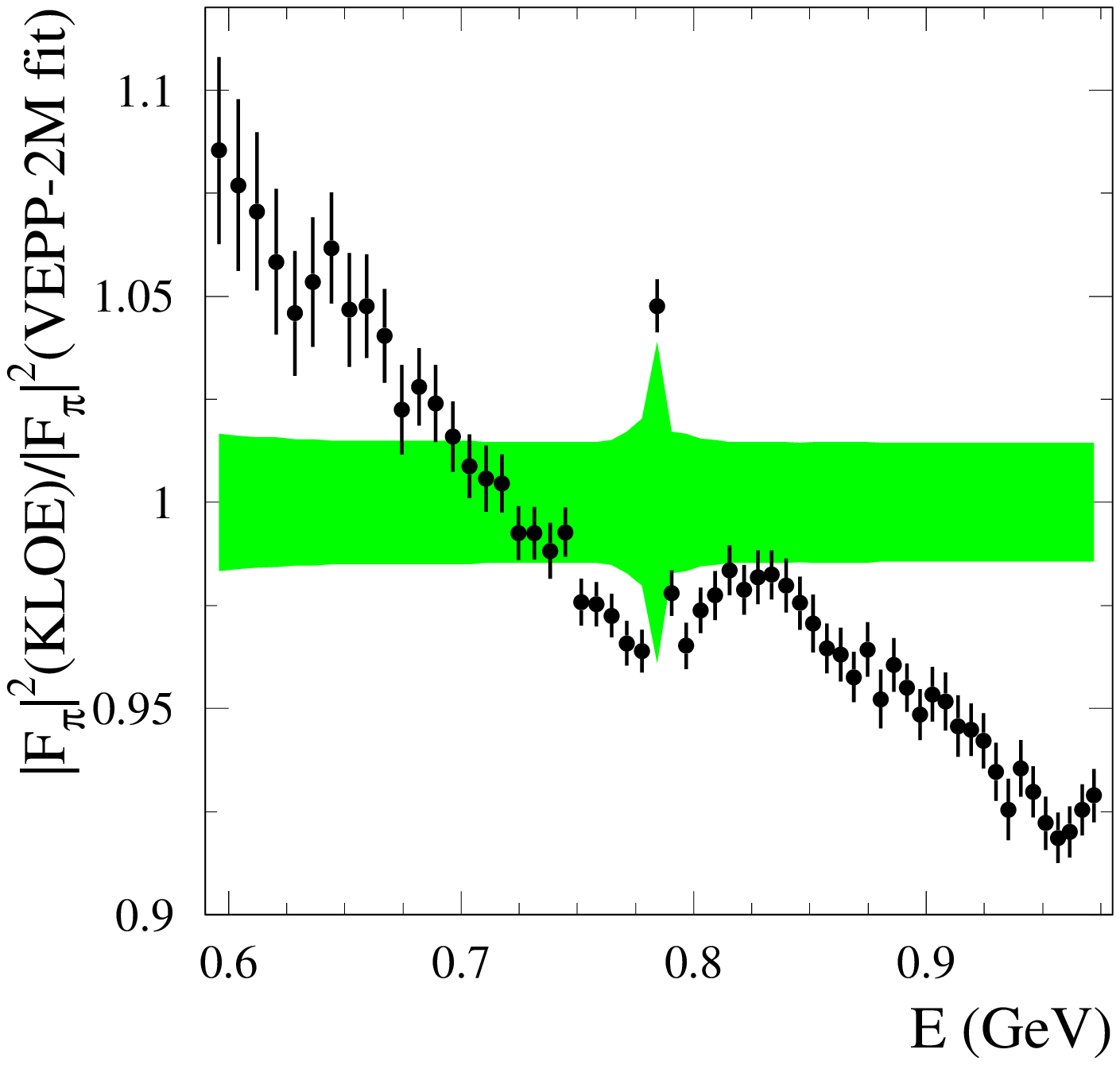}
\caption{Left: Comparison of pion form factors measured at VEPP-2M and
at KLOE. The curve is the fit to VEPP-2M data. Right: Ratio of the KLOE
pion form factor to the fit to VEPP-2M data. The shaded band shows 
joint systematic error of the form factor measurements.
\label{kloevsvepp}}
\end{figure*}

High statistics measurement of the $e^+e^-\to \pi^+\pi^-$ cross section
in the energy range 0.59--0.97 was performed by KLOE~\cite{2pi_kloe_1} 
using ISR technique on data collected in 2001.
The ISR photon was required to be emitted with $\theta_\gamma < 15^\circ$
and $\theta_\gamma > 165^\circ$. This requirement minimizes the contribution
the final state radiation which is significant at $\sqrt{s}=1.02$ GeV.
The $2\pi$ contribution into $a_\mu^{\rm had}$ evaluated by KLOE from
the measured $\pi^+\pi^-$ mass spectrum
agrees with the corresponding values calculated on CMD-2 and SND data
(see Table~\ref{tab2}). There is, however, a significant discrepancy between
the energy dependencies of pion form factors measured at VEPP-2M and
in KLOE experiment (Fig.\ref{kloevsvepp}).
\begin{table}
\caption{\label{tab2} Comparison of the $2\pi$ contributions into
$a_\mu^{\rm had}$ (in units of $10^{-10}$) calculated using
CMD-2, SND, and KLOE data from the energy range 0.630--0.958 GeV. 
The first error is statistical, the second is systematic.}
\begin{tabular}{lc}
\hline
experiment          &  $a_\mu^{\pi\pi}$  \\
\hline
CMD-2, 94-95~\cite{2pi_cmd_1}        &  $362.1 \pm 2.4 \pm 2.2$ \\
CMD-2, 98~\cite{2pi_cmd_2}           &  $361.5 \pm 1.7 \pm 2.9$ \\
SND~\cite{2pi_snd_1}                 &  $361.0 \pm 2.0 \pm 4.7$ \\
KLOE, 01~\cite{2pi_kloe_1}           &  $357.2 \pm 0.5 \pm 4.6$ \\
KLOE, 02~\cite{2pi_kloe_2}           &  $355.5 \pm 0.5 \pm 3.6$ \\
\hline
\end{tabular}
\end{table}

\begin{figure*}
\begin{minipage}[t]{0.48\textwidth}
\includegraphics[width=.98\linewidth]{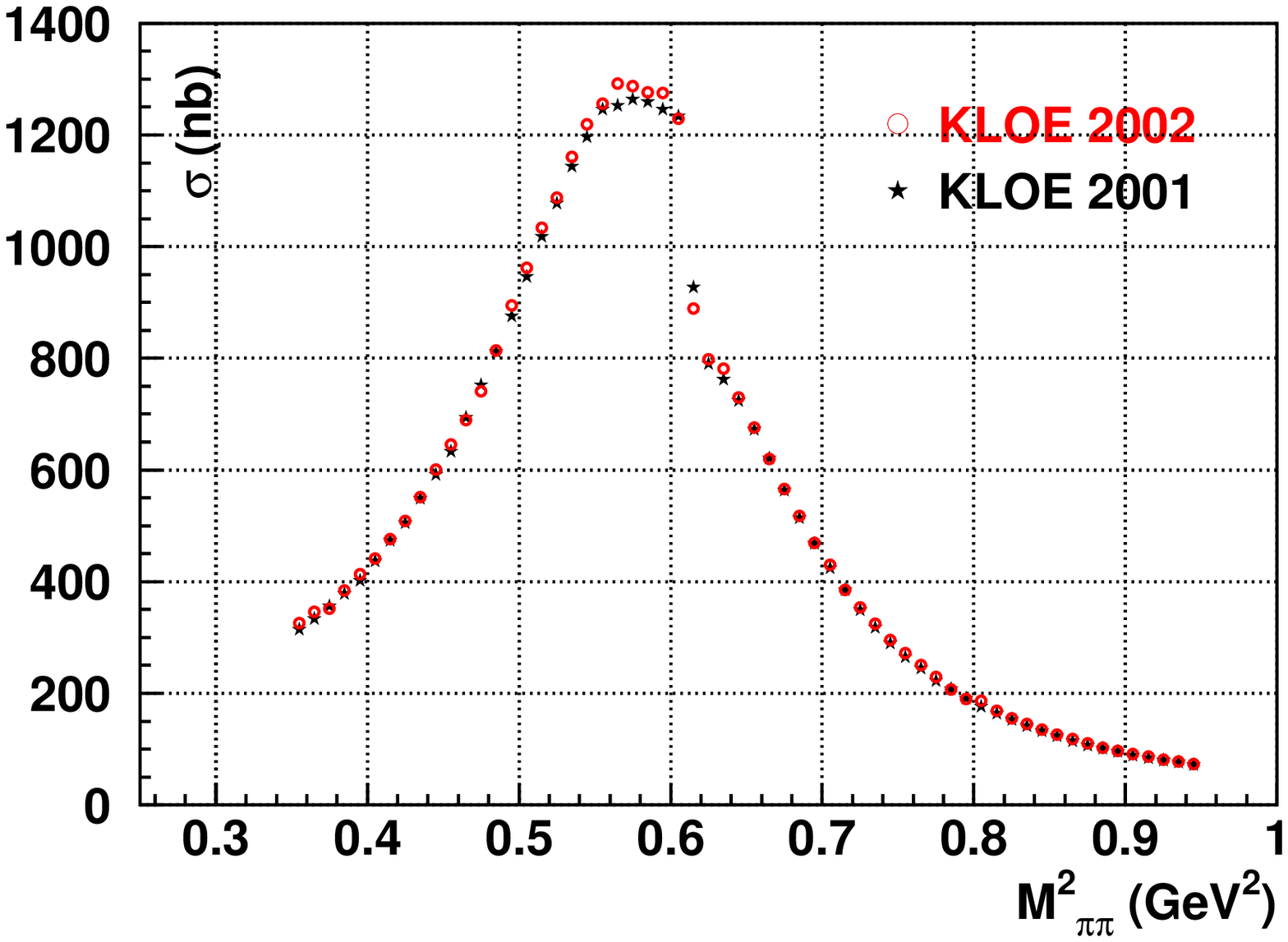}
\caption{The $e^+e^-\to \pi^+\pi^-$ cross section measured by KLOE.
The dark stars shows updated 2001 data. The preliminary 2002 data
is shown by the circles.
\label{kloe_2pi_01}}
\end{minipage}
\hfill
\begin{minipage}[t]{0.48\textwidth}
\includegraphics[width=.98\linewidth]{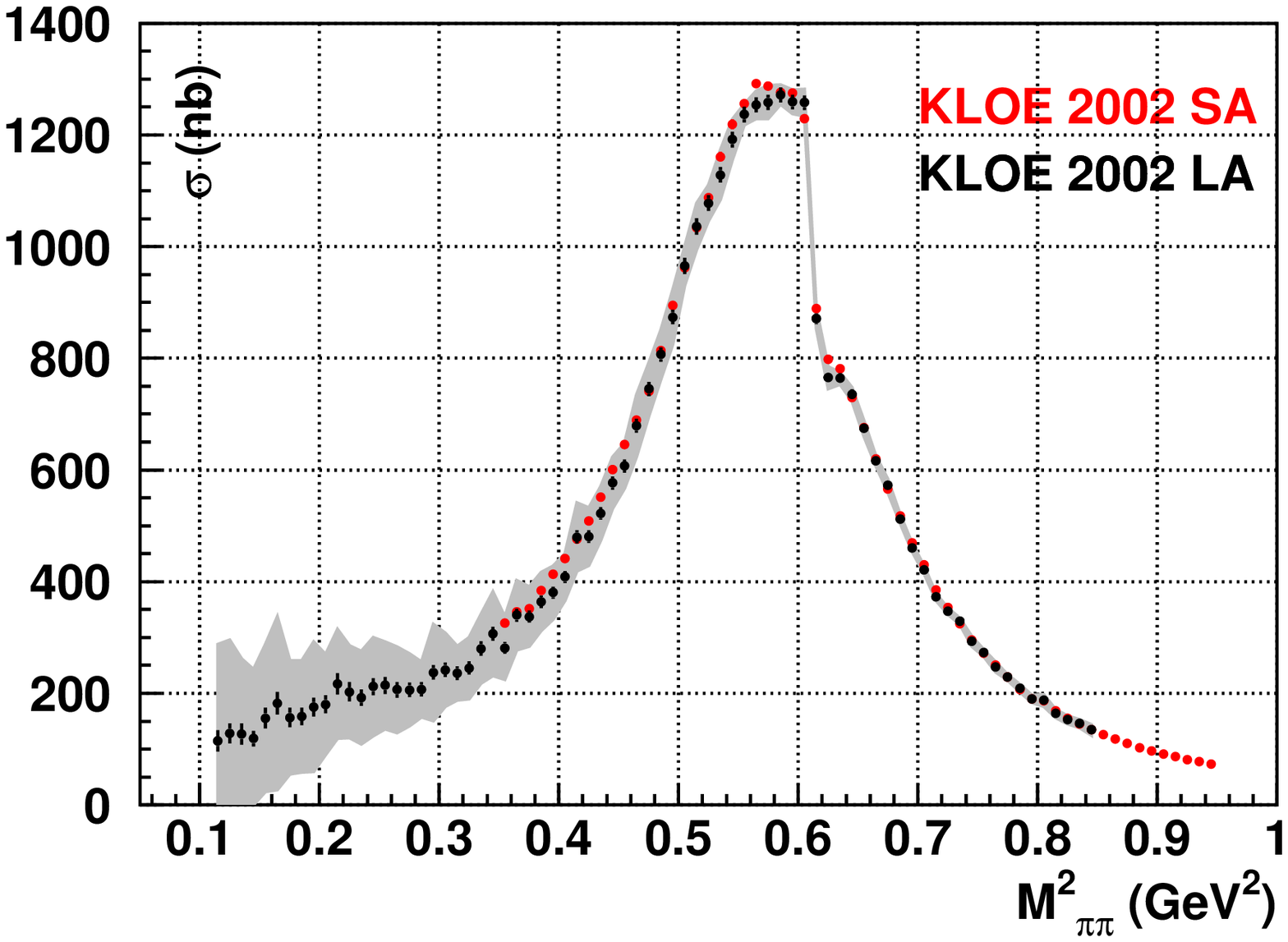}
\caption{The $e^+e^-\to \pi^+\pi^-$ cross section measured by KLOE
with the small (SA) and large (LA) angle $\gamma$ selections.
The shaded band indicates the systematic uncertainty for the LA measurement.
\label{kloe_2pi_02}}
\end{minipage}
\end{figure*}

Table~\ref{tab2} also contains the KLOE preliminary result based
on statistics collected in 2002~\cite{2pi_kloe_2}. The
new software trigger, improvement in the offline background filter, 
and more accurate calculation of the luminosity were introduced, leading
to decrease of 
systematic error. A bias in the evaluation on the trigger correction 
for the published analysis of 2001 data was found which affects at
the low $\pi^+\pi^-$ mass region. The 0.7\% decrease of the predicted
cross section for Bhabha scattering used for normalization,
also was found. The updated 2001 $e^+e^-\to \pi^+\pi^-$ cross section
and the preliminary 2002 cross section are compared in Fig.~\ref{kloe_2pi_01}.
In spite of nonstatistical difference between the two measurements seen
near the maximum of $\rho$-peak, the values of $a_\mu^{\pi\pi}$ calculated over
full mass region, $(384.4\pm0.8\pm4.9)\times10^{-10}$ for 2001 data and
$(386.3\pm0.6\pm3.9)\times10^{-10}$ for 2002 data, are in good agreement.
The KLOE also presents the measurement of $e^+e^-\to \pi^+\pi^-$ cross
section with the ISR photon detected at large angle 
($50^\circ < \theta_\gamma < 130^\circ$). In this case the threshold mass
region becomes accessible. Comparison of the small angle and large angle
measurements are demonstrated in Fig.~\ref{kloe_2pi_02}.
The systematic uncertainty for the large angle measurement is shown by the
shaded band. The main source of the uncertainty is subtraction of 
the final state radiation background. The large angle and small angle
measurements agree within the systematic errors.

The good cross check
for the ISR measurement of $e^+e^-\to \pi^+\pi^-$ is simultaneous selection
of $e^+e^-\to \mu^+\mu^-\gamma$ events. This allows to test ISR method
using the QED process with known cross section. Taking
$\pi^+\pi^-/\mu^+\mu^-$ ratio reduces the systematic uncertainty,
in particular, due to radiative effects. KLOE plans to measure
$\pi^+\pi^-/\mu^+\mu^-$ ratio for the final result.

Other KLOE result is the measurement of the $e^+e^-\to \omega\pi^0$ cross 
section in the vicinity of $\phi$-resonance~\cite{kloe_ompi}. In this region
two process, nonresonant $\omega\pi^0$ production and $\phi\to\omega\pi^0$ 
decay, determine the total $\omega\pi^0$ cross section. The decay reveals 
itself as interference pattern in the cross section energy dependence.
The process was studied in the two final states: $\pi^+\pi^-\pi^0\pi^0$ and
$\pi^0\pi^0\gamma$. In both modes the real and imaginary parts of the relative
interference amplitude were measured. The obtained interference parameters and 
$\phi\to\omega\pi^0$ branching fraction agree with the results of the previous
SND measurement~\cite{snd_ompi} but have better accuracy. From the ratio of the
$e^+e^-\to \omega\pi^0$ cross sections measured in the two different final
states, the 
ratio of the $\omega$ branching fractions was obtained: 
$B(\omega\to\pi^0\gamma)/B(\omega\to 3\pi)=0.0934\pm0.0021$.
The KLOE result has accuracy better than the world average 
value ($0.0999\pm0.0025$) and differs from it by 2 standard deviations.
The inclusion of the KLOE result into the global fit of the $\omega$ parameters
will lead to growth of the $\omega$ leptonic width.
\section{BABAR ISR measurements}
\begin{figure*}
\includegraphics[width=.9\linewidth]{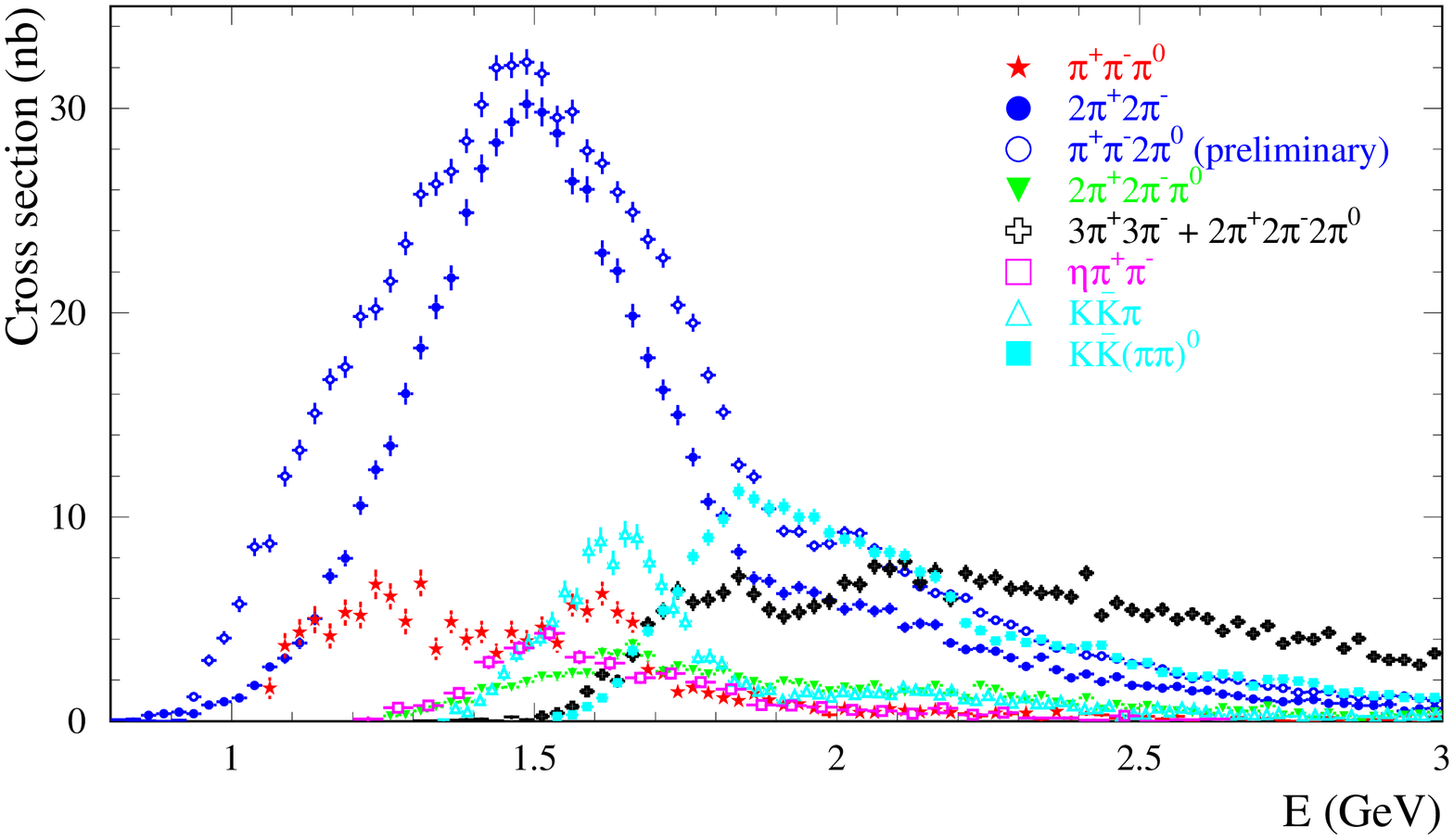}
\caption{The hadronic cross sections measured by BABAR using ISR technique.
\label{babar}}
\end{figure*}

The BABAR ISR program includes the study of all significant hadronic processes 
$e^+e^-\to f$ from threshold to c.m. energy about 4.5 GeV. One of the purposes
of this program is measurement of $R$ in the 1--2 GeV energy range with 
improved accuracy. The current status of BABAR measurements in this mass range
is demonstrated in Fig.~\ref{babar}. To obtain the total hadronic cross section
the $\pi^+\pi^-$, $\pi^+\pi^-3\pi^0$, $\pi^+\pi^-4\pi^0$, $K^+K^-$, $K_SK_L$,
$K_SK_L\pi\pi$, $K_SK^+\pi^-\pi^0$ processes must be measured additionally.
The final states $\pi^+\pi^-$, $\pi^+\pi^-3\pi^0$, $K^+K^-$, and 
$K_SK^+\pi^-\pi^0$ are currently under study. Here we discuss the latest 
BABAR results on $\pi^+\pi^-2\pi^0$, $2\pi^+2\pi^-\pi^0$, $2\pi^+2\pi^-\eta$,
$K^+K^-\pi^0$, $K_SK^+\pi^-$, $K^+K^-\pi^+\pi^-\pi^0$ modes and measurement
of strange baryon form factors.

\begin{figure*}
\begin{minipage}[t]{0.48\textwidth}
\includegraphics[width=.98\linewidth]{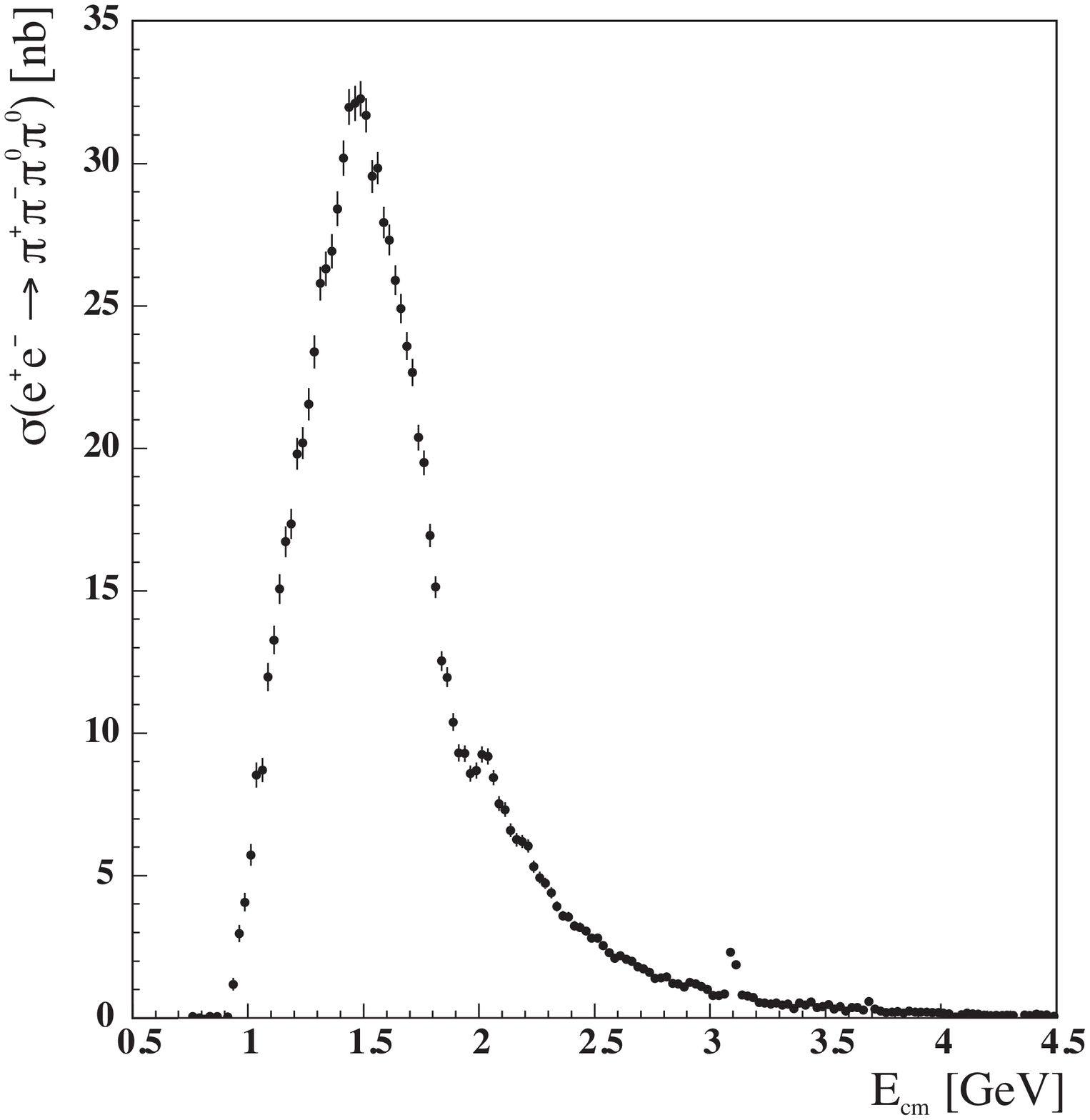}
\caption{The $e^+e^-\to \pi^+\pi^-2\pi^0$ cross section measured by BABAR.
\label{4pi_1}}
\end{minipage}
\hfill
\begin{minipage}[t]{0.48\textwidth}
\includegraphics[width=.98\linewidth]{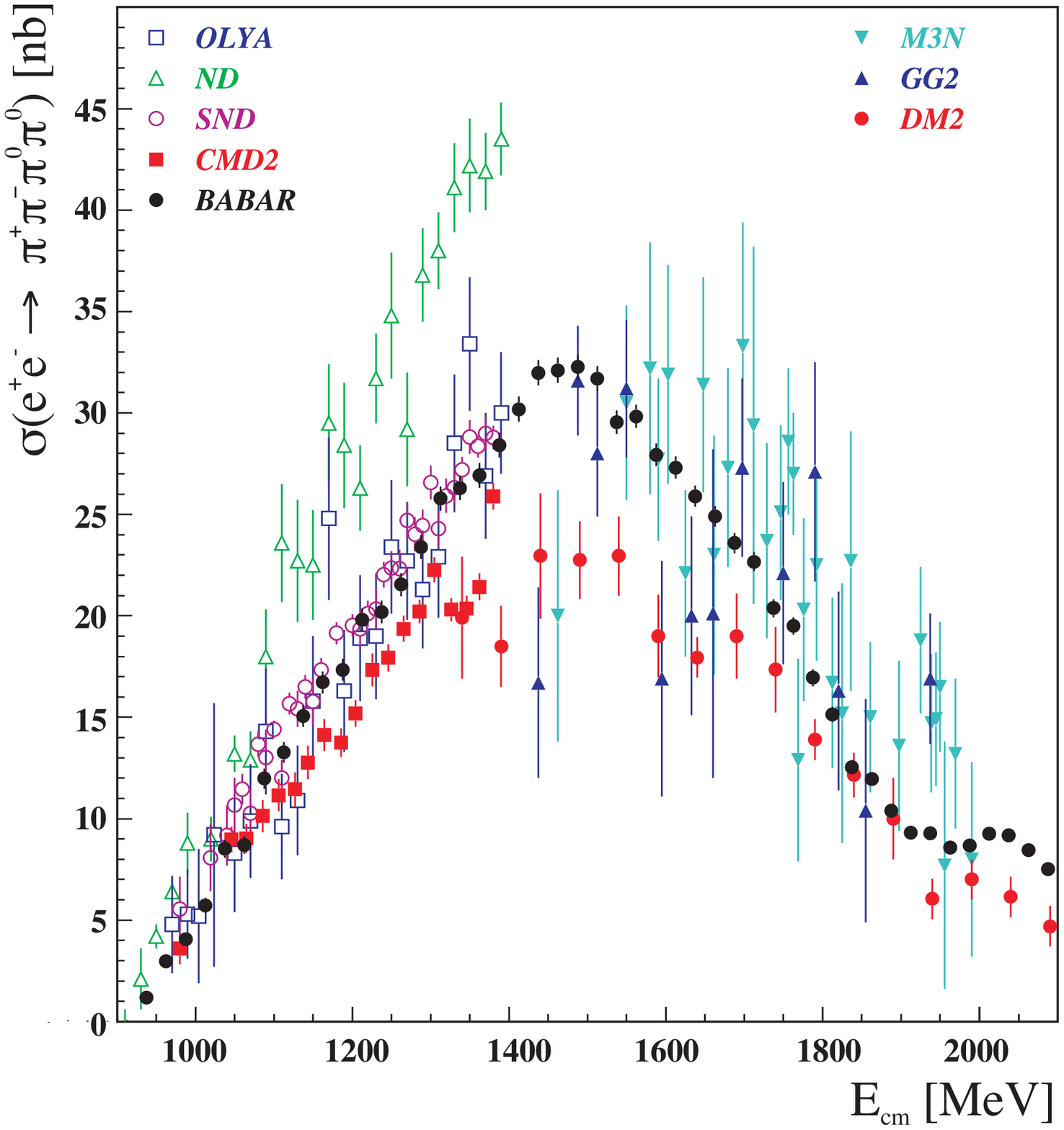}
\caption{Comparison the BABAR data on $e^+e^-\to \pi^+\pi^-2\pi^0$ cross 
section with the data of other experiments.
\label{4pi_2}}
\end{minipage}
\end{figure*}

{\bf\boldmath $e^+e^-\to \pi^+\pi^-2\pi^0$.}
This channel is very important for calculation of $a_\mu^{\rm had}$ and
spectroscopy of the excited $\rho$-like states. 
The BABAR preliminary results on $e^+e^-\to \pi^+\pi^-2\pi^0$ cross section
measurement is presented in Fig.~\ref{4pi_1}. The current systematic error
of the measurement varies from 8\% in the cross section peak to 14\% at
4.5 GeV. The comparison with existing data is shown in Fig.~\ref{4pi_2}.
In the energy range below 1.4 GeV the the BABAR results are in good agreement
with SND data. For higher energies the accuracy of the cross section is
significantly improved. The BABAR data exceed the cross section
measured by DM2 by about 25\%. From the analysis of $3\pi$ and $2\pi$ mass distributions
it was found that the $\omega\pi^0$, $a_1\pi$ and $\rho^+\rho^-$ are dominant
intermediate states for $e^+e^-\to \pi^+\pi^-2\pi^0$ process. The 
$\rho^0 f_0(980)$ contribution is also seen. The 
$e^+e^-\to\rho^+\rho^-$ process was observed for the first time, 
its contribution was found be surprisingly large, about 30\% of the total 
$e^+e^-\to\pi^+\pi^-2\pi^0$ cross section at $E=1.6$ GeV.

\begin{figure*}
\begin{minipage}[t]{0.48\textwidth}
\includegraphics[width=.98\linewidth]{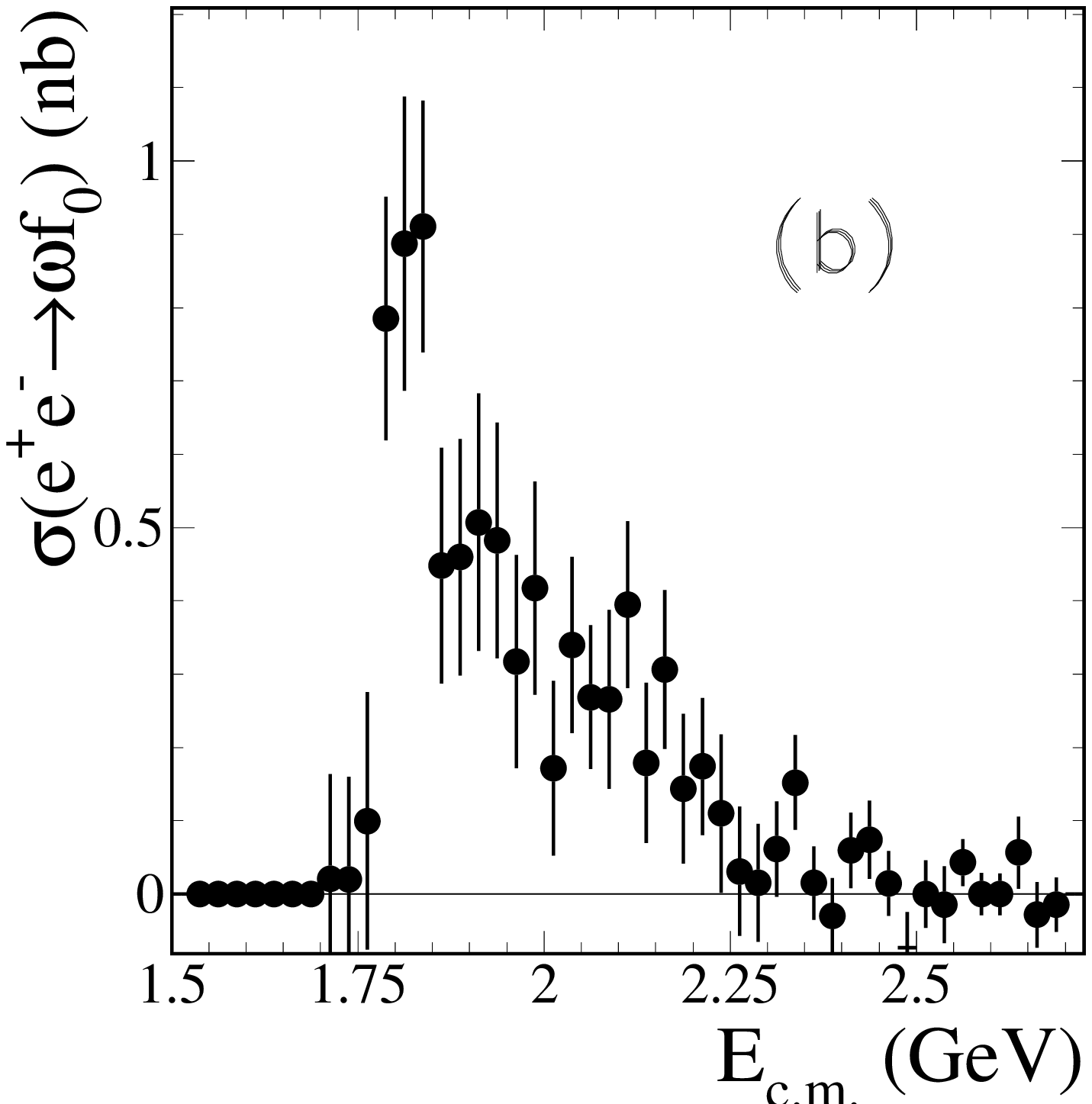}
\caption{The $e^+e^-\to \omega f_0(980)$ cross section measured by BABAR.
\label{babar_omf0}}
\end{minipage}
\hfill
\begin{minipage}[t]{0.48\textwidth}
\includegraphics[width=.98\linewidth]{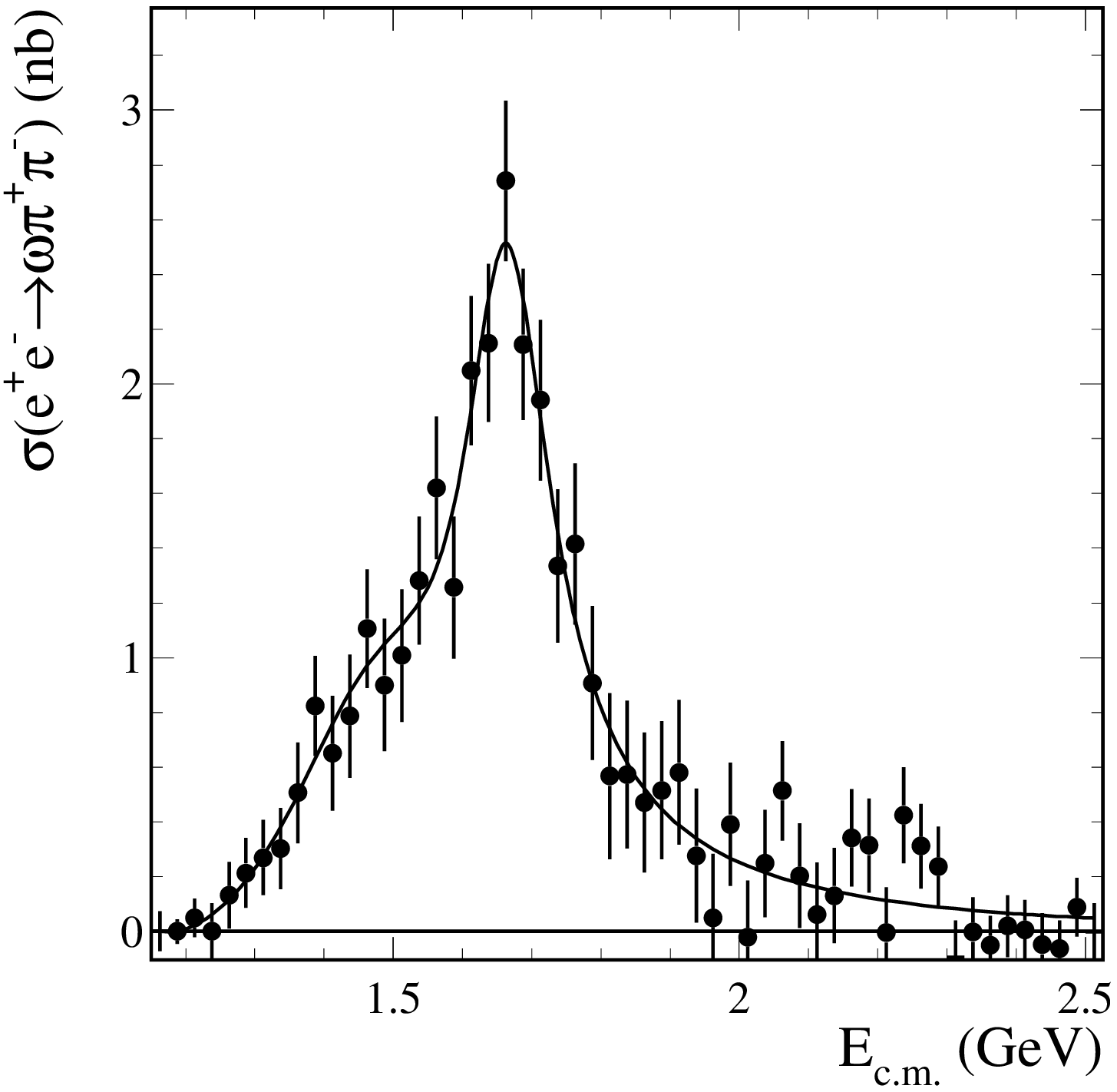}
\caption{The $e^+e^-\to \omega \pi^+\pi^-$ cross section 
with subtracted $\omega f_0(980)$ contribution. The curve
is the fit result.
\label{babar_om2pi}}
\end{minipage}
\end{figure*}

{\bf\boldmath $e^+e^-\to 2\pi^+2\pi^-\pi^0$~\cite{babar_5pi}.}
The measured cross section for $e^+e^-\to 2\pi^+2\pi^-\pi^0$ is
shown in Fig.~\ref{babar}. The systematic error of the measurement
is about 7\%. In $\pi^+\pi^-\pi^0$ mass spectrum the $\omega$ and
$\eta$ peaks were observed corresponding the $\omega\pi^+\pi^-$
and $\eta\pi^+\pi^-$ intermediate states. Non-$\omega\pi^+\pi^-$ and
non-$\eta\pi^+\pi^-$ events have $\rho X = \rho^0 3\pi + \rho^\pm 3\pi$
structure, where $X$ may be $\pi(1300)$ or $a_1(1360)$. The cross
sections for all three selected components were measured. The 
$\eta\pi^+\pi^-$ and $\omega\pi^+\pi^-$ cross sections agree with
previous experiments but have significantly better accuracy.  
The $e^+e^-\to \eta\pi^+\pi^-$ cross
section is shown in Fig.~\ref{babar}. A part of $\omega\pi^+\pi^-$
events contain $f_0(980)$ meson. The cross sections for the $\omega f_0$ and
non-$\omega f_0$ components of the $\omega\pi^+\pi^-$ final state is
shown in Figs.~\ref{babar_omf0} and ~\ref{babar_om2pi}. The latter
cross section was fitted by the sum of the contributions of
$\omega$-like resonances. From the fit, the parameters
$\omega^{\prime\prime}$ meson, $m=1.667\pm 0.014$ GeV and
$\Gamma = 0.22\pm0.03$, were determined.   

\begin{figure*}
\includegraphics[width=.32\linewidth]{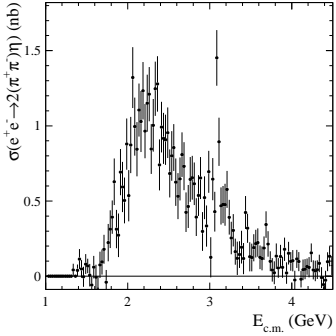}
\includegraphics[width=.32\linewidth]{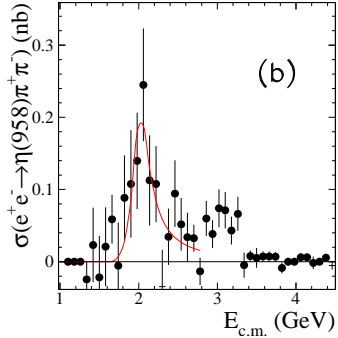}
\hfill
\includegraphics[width=.32\linewidth]{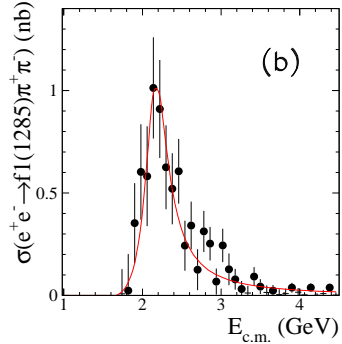}
\caption{From left to right: The $e^+e^-\to 2\pi^+2\pi^-\eta$,
$e^+e^-\to \eta^\prime\rho(770)$ and $e^+e^-\to f_1(1285)\rho(770)$ 
cross sections measured by BABAR. The curves are single Breit-Wigner
fits.
\label{babar_4pieta}}
\end{figure*}

{\bf\boldmath $e^+e^-\to 2\pi^+2\pi^-\eta$~\cite{babar_5pi}.}
The $2\pi^+2\pi^-\eta$ final state was studied for the first time.
Three intermediate states, $\eta\rho(1450)\to\eta 4\pi$,
$\eta^\prime\rho(770)$, and $f_1(1285)\rho(770)$, were found to
contribute to $2\pi^+2\pi^-\eta$. The $e^+e^-\to 2\pi^+2\pi^-\eta$
total cross section, and the cross sections for $\eta^\prime\rho(770)$ and
$f_1(1285)\rho(770)$ final states are shown in Fig.~\ref{babar_4pieta}. The
two latter cross sections are described by the single Breit-Wigner function
with parameters: $m=1.99\pm0.08$ GeV, $\Gamma=0.31\pm0.14$ GeV for
$\eta^\prime\rho(770)$ and $m=2.15\pm0.07$ GeV, $\Gamma=0.35\pm0.07$ GeV for
$f_1(1285)\rho(770)$. Both sets of parameters are close to those listed in
the PDG tables~\cite{pdg} for the $\rho(2150)$.

\begin{figure*}
\includegraphics[width=.32\linewidth]{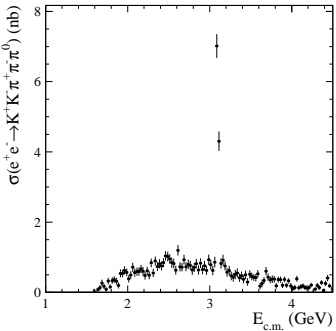}
\includegraphics[width=.32\linewidth]{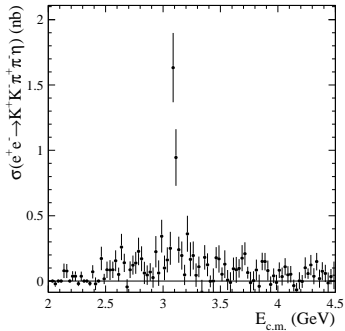}
\hfill
\includegraphics[width=.32\linewidth]{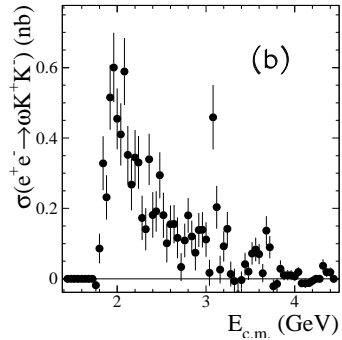}
\caption{From left to right: The $e^+e^-\to K^+K^-\pi^+\pi^-\pi^0$,
$e^+e^-\to K^+K^-\pi^+\pi^-\eta$ and $e^+e^-\to \omega K^+K^-$ 
cross sections measured by BABAR.
\label{babar_2k3pi}}
\end{figure*}

{\bf\boldmath $e^+e^-\to K^+K^-\pi^+\pi^-\pi^0$ and 
$K^+K^-\pi^+\pi^-\eta$~\cite{babar_5pi}.}
The $e^+e^-\to K^+K^-\pi^+\pi^-\pi^0$ and $e^+e^-\to K^+K^-\pi^+\pi^-\eta$
cross sections measured for the first time are shown in Fig.~\ref{babar_2k3pi}.
In $3\pi$ mass spectrum for $K^+K^-\pi^+\pi^-\pi^0$ channel the $\eta$ and 
$\omega$ peaks are observed corresponding the $\omega K^+K^-$ and $\eta\phi$
intermediate states. The cross section for $e^+e^-\to \omega K^+K^-$ is shown
in Fig.~\ref{babar_2k3pi}. The $e^+e^-\to\eta\phi$ process will be discussed
in the next subsection.

\begin{figure*}
\includegraphics[width=.32\linewidth]{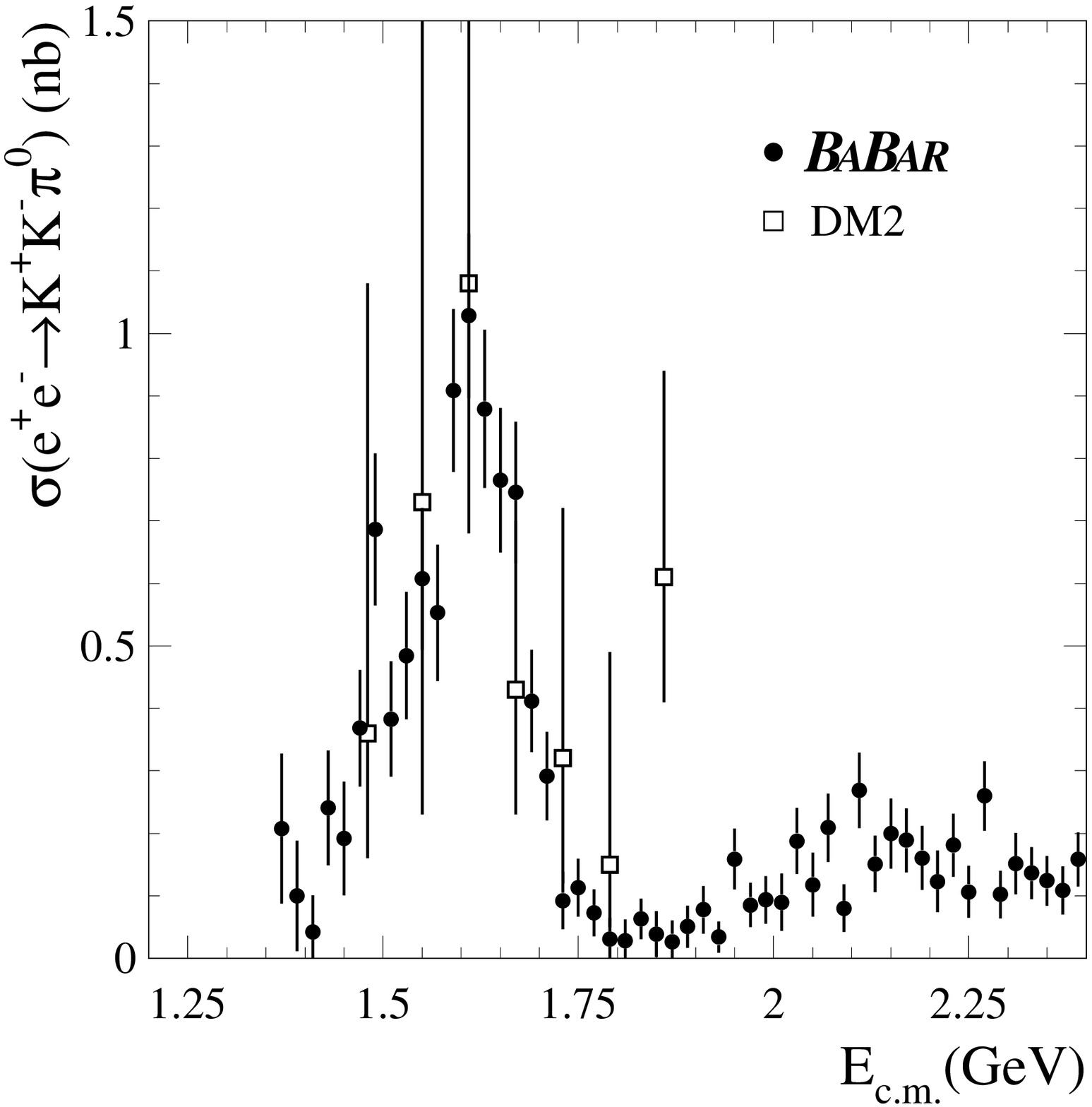}
\includegraphics[width=.32\linewidth]{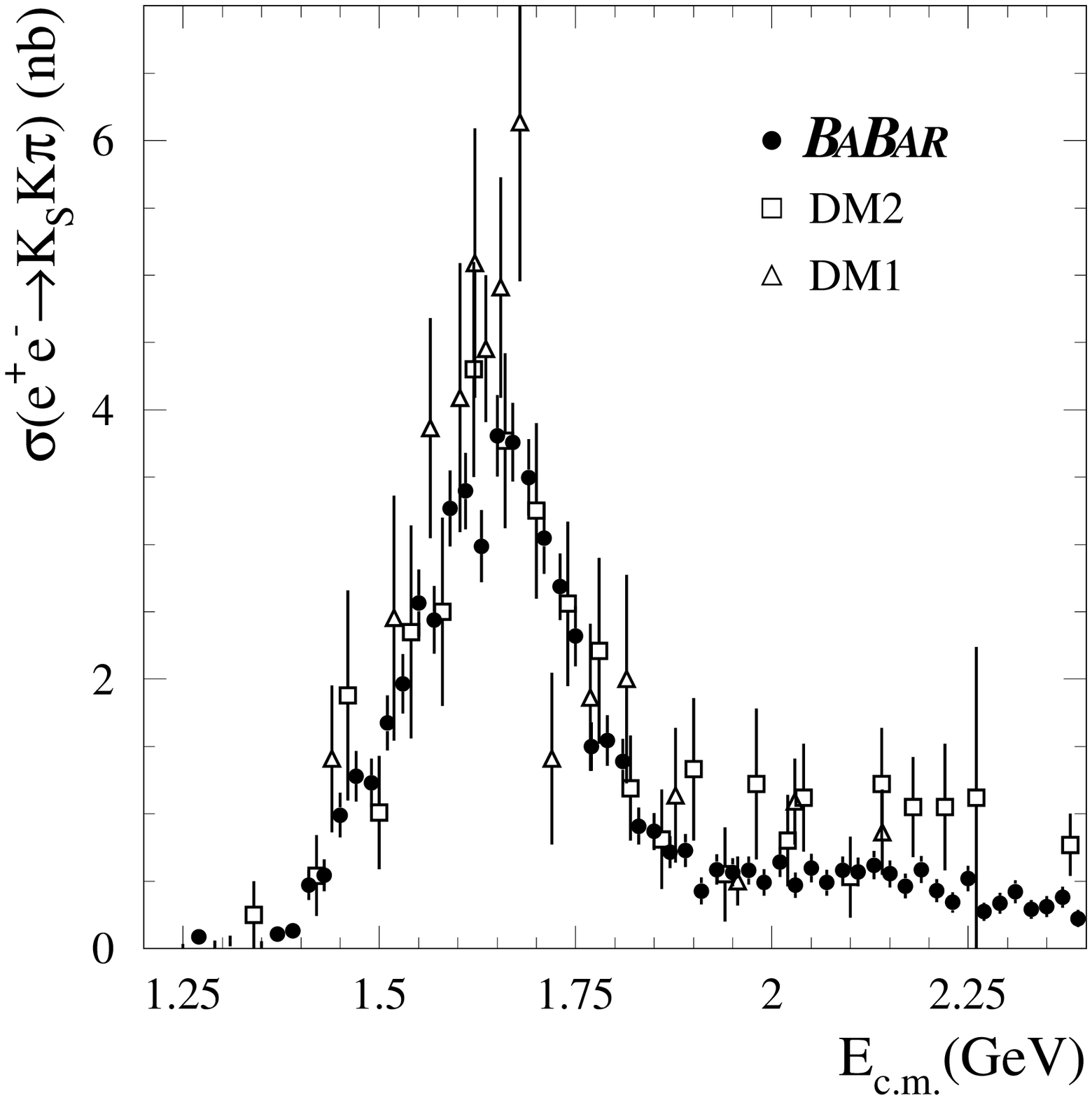}
\hfill
\includegraphics[width=.32\linewidth]{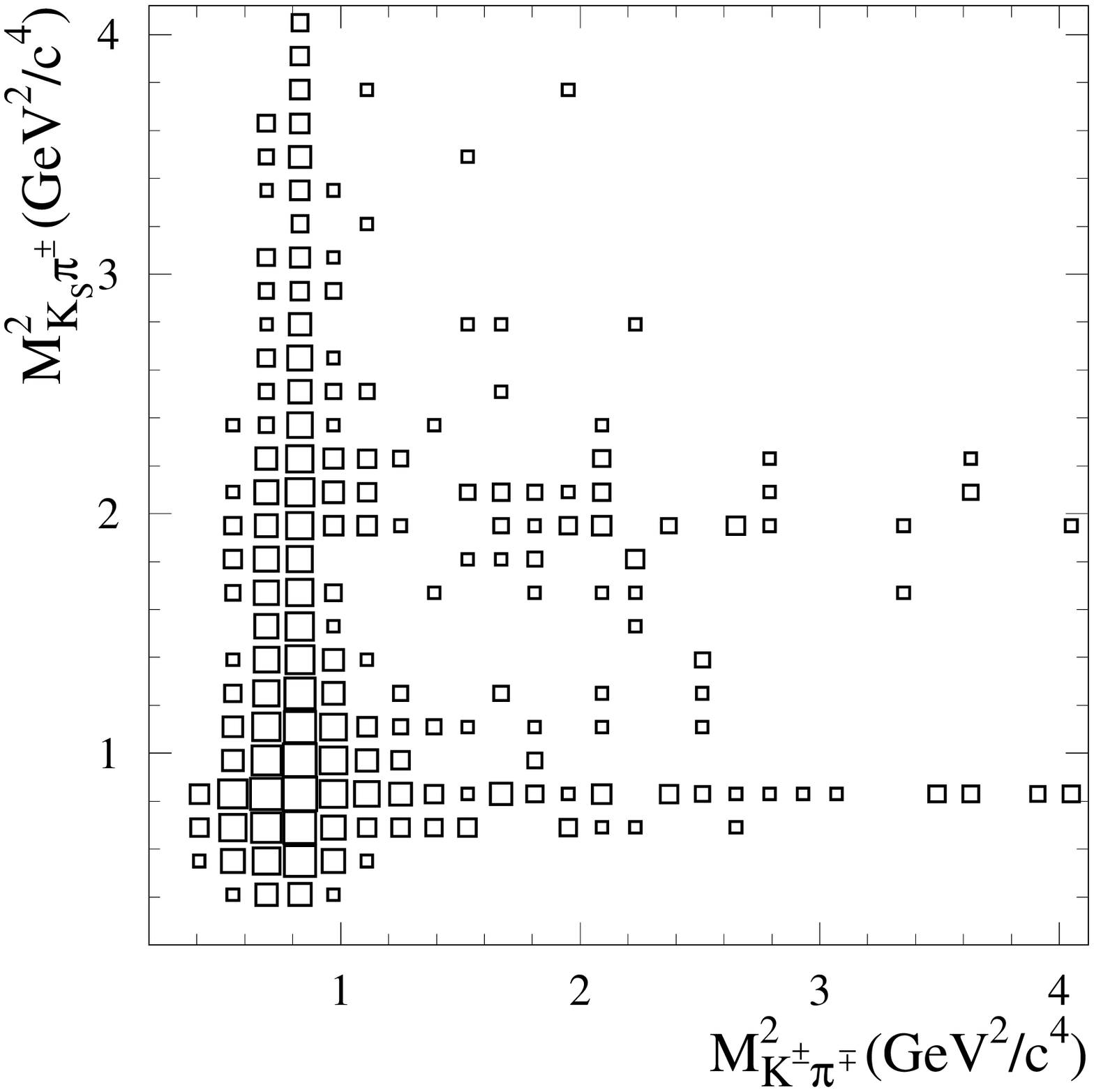}
\caption{The $e^+e^-\to K^+K^-\pi^0$ (left) and $e^+e^-\to K_S K\pi$ (middle) 
cross sections measured by BABAR in comparison with results of previous 
measurements. Right: The Dalitz plot for $e^+e^-\to K_S K\pi$.
\label{babar_kkpi}}
\end{figure*}

\begin{figure*}
\includegraphics[width=.32\linewidth]{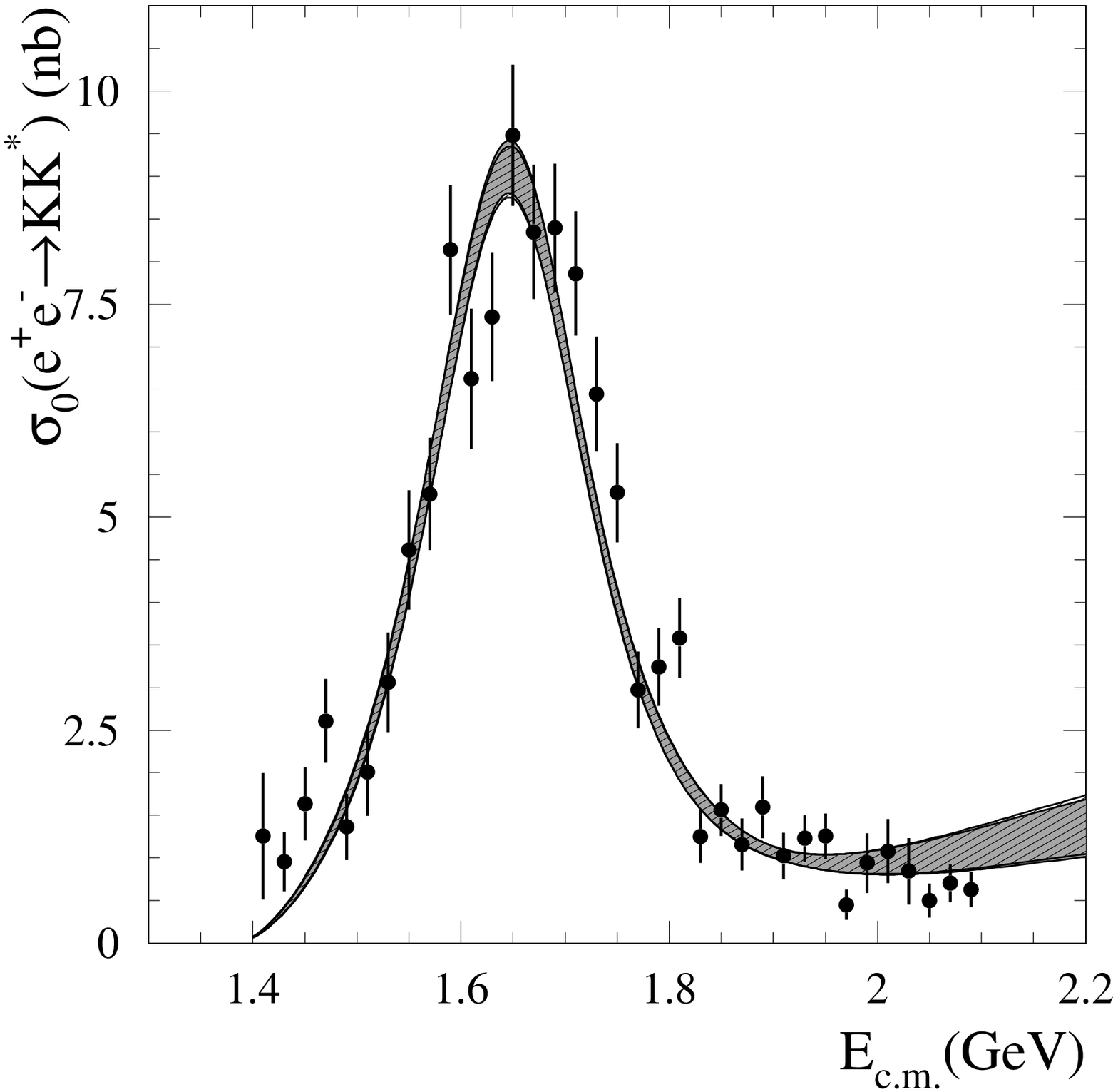}
\includegraphics[width=.32\linewidth]{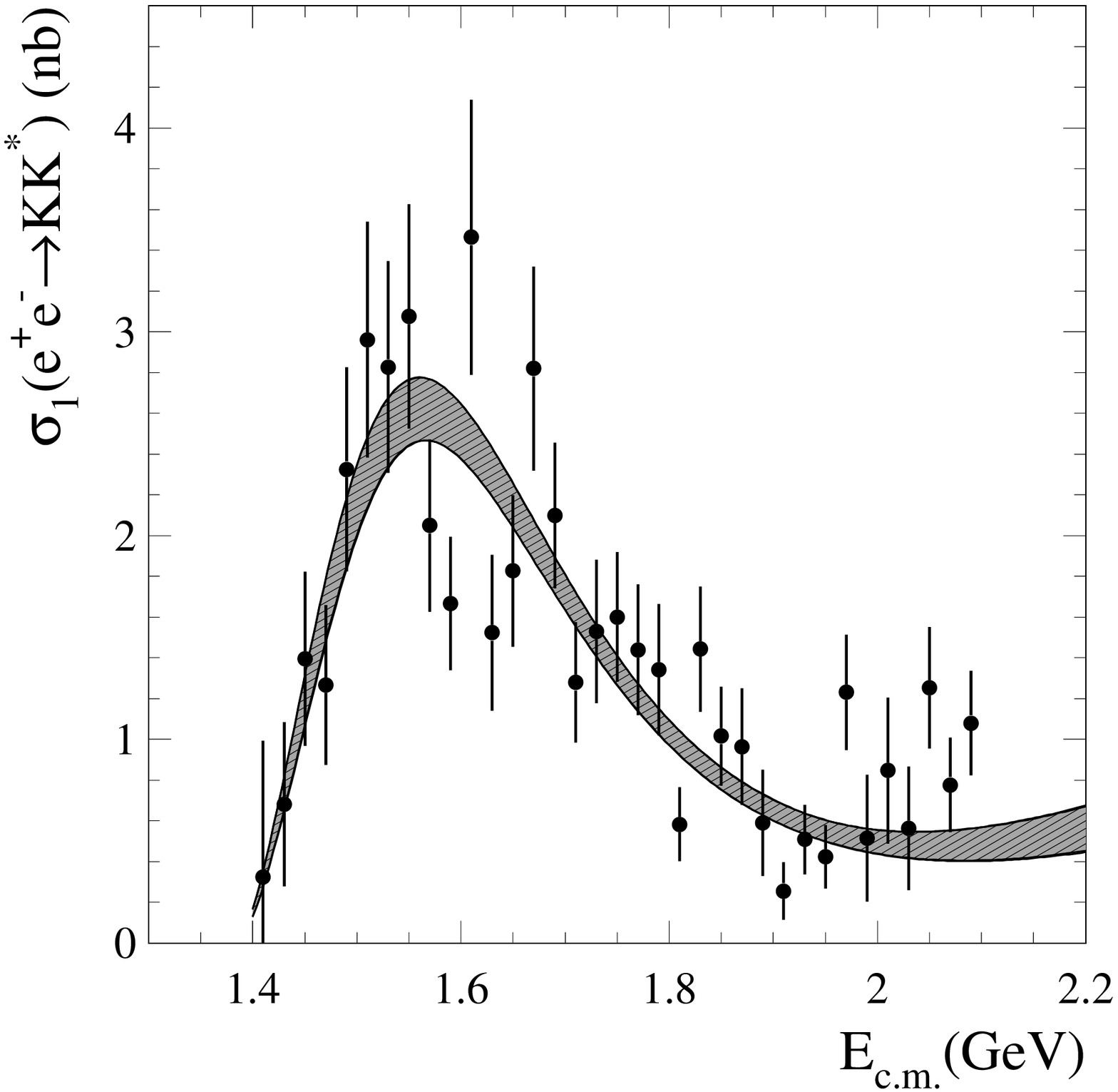}
\hfill
\includegraphics[width=.32\linewidth]{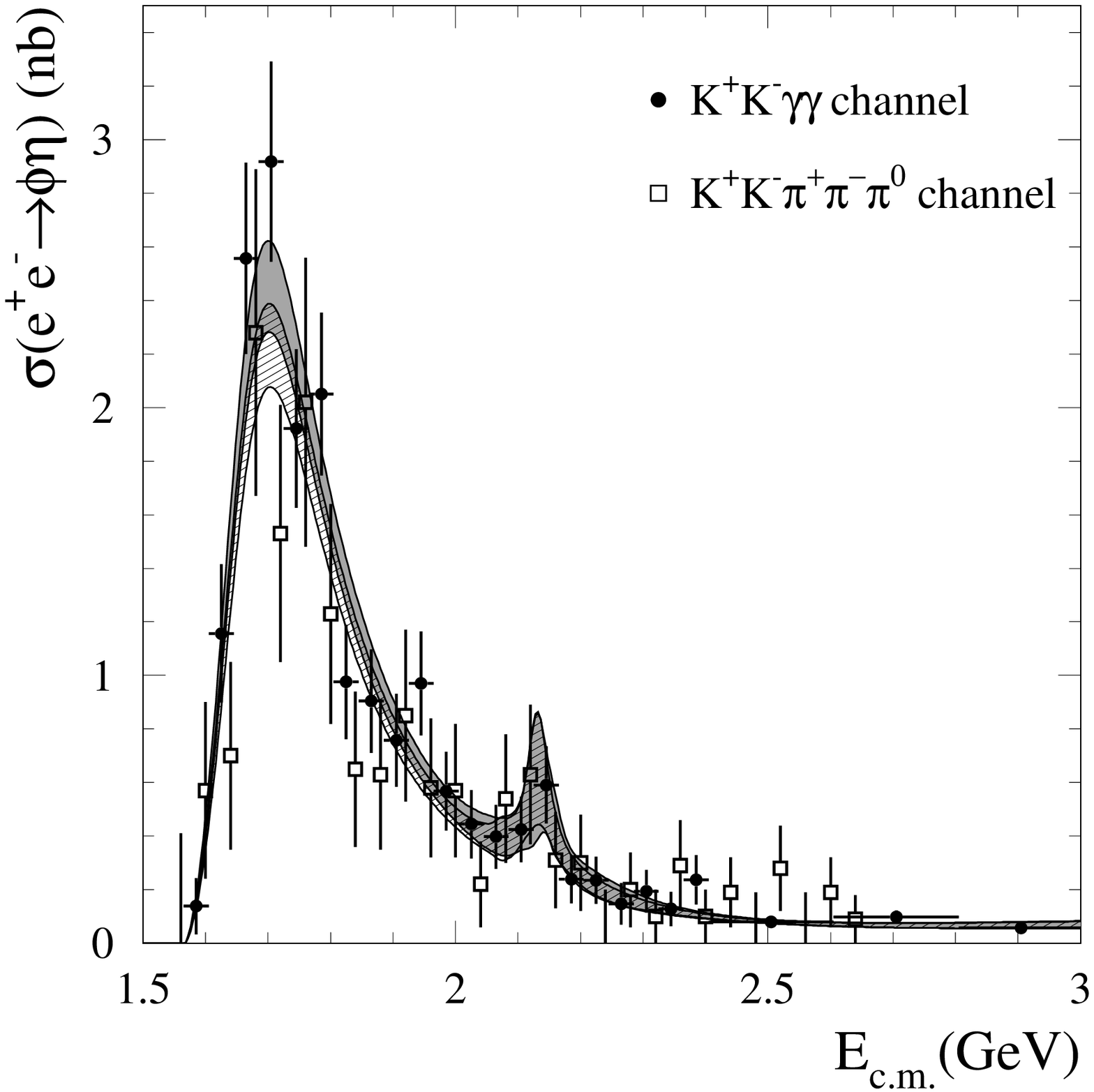}
\caption{The isoscalar (left) and isovector (middle) $e^+e^-\to KK^\ast(892)$ 
cross sections and $e^+e^-\to \phi\eta$ (right) cross section measured by
BABAR. The bands represent the fits.
\label{babar_phieta}}
\end{figure*}

{\bf\boldmath $e^+e^-\to K^+K^-\pi^0$, $K_SK^\pm\pi^\mp$, 
$\phi\pi^0$, $\phi\eta$.}
The $K\bar{K}\pi$ is the dominant decay mode of the $\phi(1680)$ meson. Therefore
the study of $e^+e^-\to K\bar{K}\pi$ process is very important for 
determination of its parameters. This process was studied at BABAR in the two
final states, $K^+K^-\pi^0$ and $K_SK^\pm\pi^\mp$. The measured cross 
sections for both channels are shown in Fig.~\ref{babar_kkpi}. The
BABAR results are in reasonable agreement with the results of the previous
measurements by DM1 and DM2 but have much smaller statistical and systematic
errors. The systematic uncertainty of the BABAR measurement is about 5\%.
Fig.~\ref{babar_kkpi}(right) shows the Dalitz plot for $K_SK^\pm\pi^\mp$ final
state. It is seen that the $KK^\ast(892)$ and $KK_2^\ast(1430)$ intermediate
state give the main contributions to the $K\bar{K}\pi$ production.
For $K_SK^\pm\pi^\mp$ final state both neutral $K^0K^{\ast 0}$ and charged
$K^\pm K^{\ast \mp}$ intermediate states are possible. Since the 
${K}^0K^{\ast 0}$ and 
$K^\pm K^{\ast \mp}$ amplitudes are the sum and the difference of the
isovector and isoscalar amplitudes, the Dalitz plot population strongly 
depends on isospin composition. From the Dalitz plot analysis the moduli and relative
phase of the isoscalar and isovector amplitudes both for $KK^\ast(892)$ and
$KK_2^\ast(1430)$ final states were determined. The obtained isoscalar and
isovector $e^+e^-\to KK^\ast(892)$ cross sections is shown in 
Fig.~\ref{babar_phieta}.

The $e^+e^-\to \phi\eta$ process was studied by BABAR in the
$K^+K^- 2\gamma$ and $K^+K^-3\pi$ final states. The resulted cross section is
shown in Fig.~\ref{babar_phieta}(right). The $e^+e^-\to \phi\eta$ is the best
channel for study of excited $\phi$ state. The contributions of $\omega$-like
states are suppressed by OZI rule. The global fit of the $e^+e^-\to
\phi\eta$ and $e^+e^-\to K^+K^-\pi^0$ cross sections, isovector and isoscalar
$KK^\ast(892)$ amplitudes, and their relative phase was performed to
determinate parameters of the $\phi$ and $\rho$ excitations decaying into these
final states. The fit results are shown in Fig.~\ref{babar_phieta}.
The isovector $KK^\ast(892)$ component is described by a single broad
resonance which mass is compatible with that for $\rho(1450)$~\cite{pdg}.
The obtained $\phi^\prime$ parameters are following: $m=1723\pm 24$ MeV,
$\Gamma=371\pm 90$ MeV, $\Gamma_{ee}=580\pm 60$ eV, 
$B(\phi^\prime\to \phi\eta)/B(\phi^\prime\to KK^\ast)\approx 1/3$.

In the $\phi\eta$ cross section the peak with $m=2139\pm 35$ MeV and
$\Gamma=76\pm 62$ MeV is seen with 2 sigma significance. These parameters are
close to the mass and width of the state recently observed by BABAR 
in $\phi f_0(980)$ mode~\cite{babar_phif0}.

\begin{figure*}
\begin{minipage}[t]{0.48\textwidth}
\includegraphics[width=.98\linewidth]{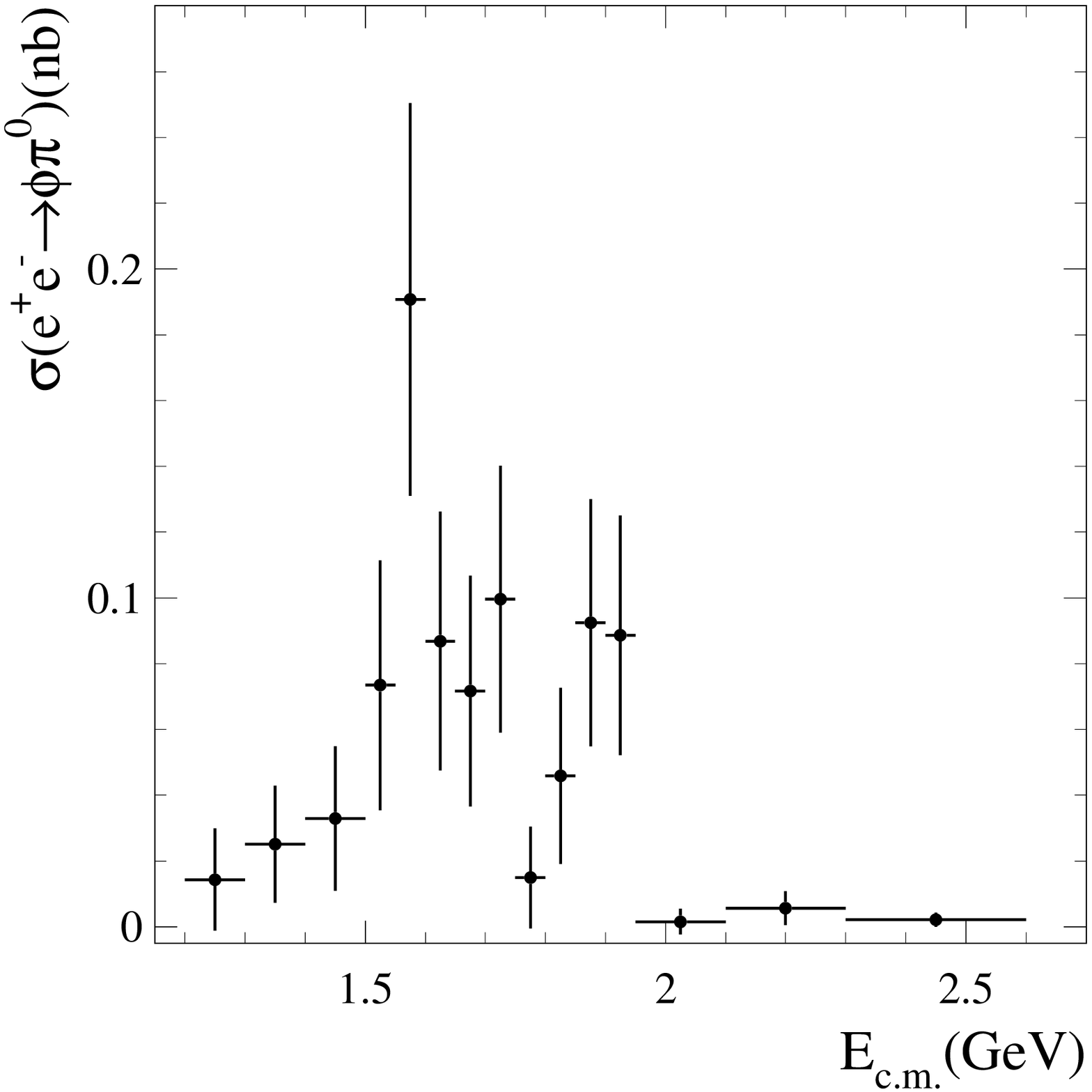}
\caption{The $e^+e^-\to \phi \pi^0$ cross section measured by BABAR.
\label{babar_phipi0}}
\end{minipage}
\hfill
\begin{minipage}[t]{0.48\textwidth}
\includegraphics[width=.98\linewidth]{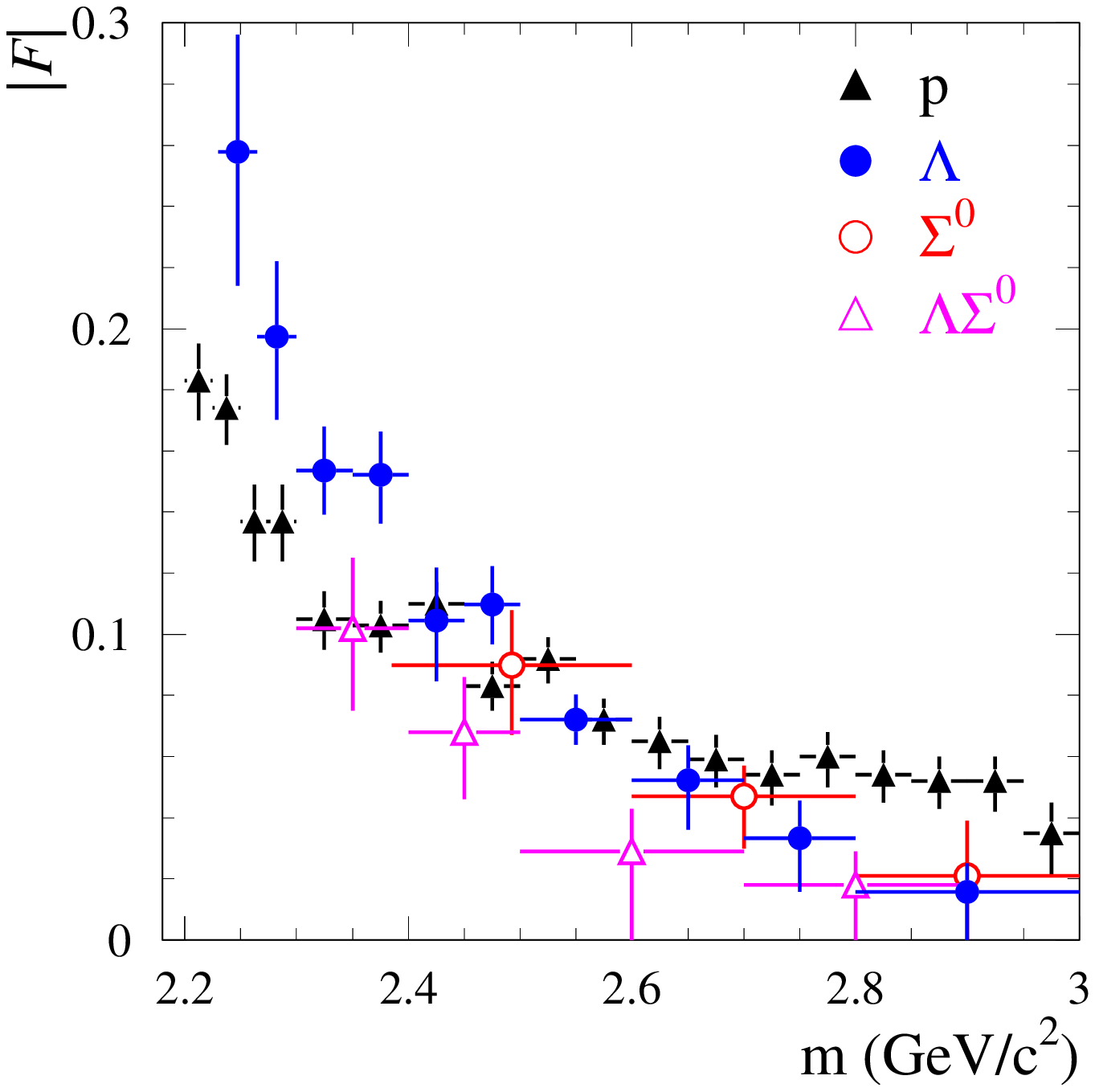}
\caption{The octet baryon form factors measured by BABAR. 
\label{babar_baryons}}
\end{minipage}
\end{figure*}

The cross section for $e^+e^-\to \phi \pi^0$ measured by BABAR in the 
$K^+K^-\pi^0$ final state is shown in Fig.~\ref{babar_phipi0}.
This reaction is suitable for search of exotic isovector resonances. For
the ordinary isovector states, $\phi \pi^0$ decay is suppressed by OZI rule.
With available limited statistics the $e^+e^-\to \phi \pi^0$ cross section
is described by the single resonance with $m=1600\pm 30$ MeV and
$\Gamma=200\pm 100$ MeV. These parameters are compatible with those for
$\rho(1700)$~\cite{pdg}.

{\bf\boldmath Baryon form factors~\cite{babar_lambda}.}
BABAR continued the study of the baryon time-like electromagnetic
form factors started by the work~\cite{babar_ppbar} on 
the $e^+e^-\to p\bar{p}$ reaction. The cross section for
$e^+e^-\to B\bar{B}$, where $B$ is a spin-1/2 baryon depends on
two form factors, magnetic ($G_M$) and electric ($G_E$).
From the measurements of the total cross section for 
$e^+e^-\to \Lambda\Lbar$, $\Lambda\Sigbar^0$, and $\Sigma^0\Sigbar^0$
reactions, the effective 
$\Lambda$, $\Lambda$-$\Sigma^0$, and $\Sigma^0$ 
form factors were extracted. These form 
factors together with the previously measured~\cite{babar_ppbar} proton
form factor are shown in Fig.~\ref{babar_baryons}. For $\Lambda\Lbar$
channel the $\Lambda$ angular distributions were studied and the ratio
of the form factors was determined for the two energy
regions: $|G_E/G_M|=1.7^{+1.0}_{-0.6}$ for 2.23--2.40 GeV and
$|G_E/G_M|=0.7\pm0.7$ for 2.4--2.8 GeV. The obtained ratios are consistent
both with $|G_E/G_M|=1$ and with the results for $e^+e^-\to p\bar{p}$,
where this ratio was found to be significantly greater than unity near
threshold. The measurement of $\Lambda$ polarization via its decay  to
$p\pi^-$ gives possibility to extract the relative phase between the electric and
magnetic form factors. The limited experimental statistics allowed to set
only very weak limits on this phase: $-0.76 < \sin{\phi} < 0.98$.

\section{Tests of CVC}
\begin{figure*}
\includegraphics[width=.7\linewidth]{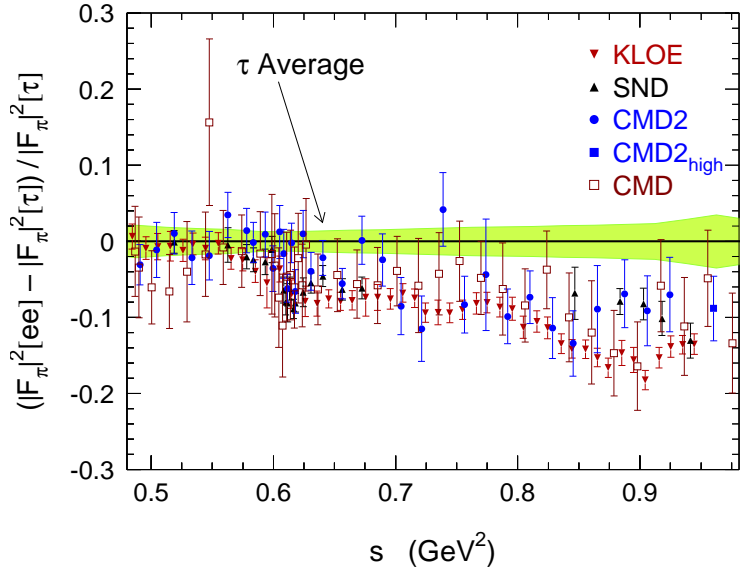}
\caption{Comparison of the pion form factors obtained from $e^+e^-$-annihilation
and from $\tau$ data. The shaded band indicates the errors of $\tau$ data.
\label{tau2pi}}
\end{figure*}

\begin{figure*}
\includegraphics[width=.4\linewidth]{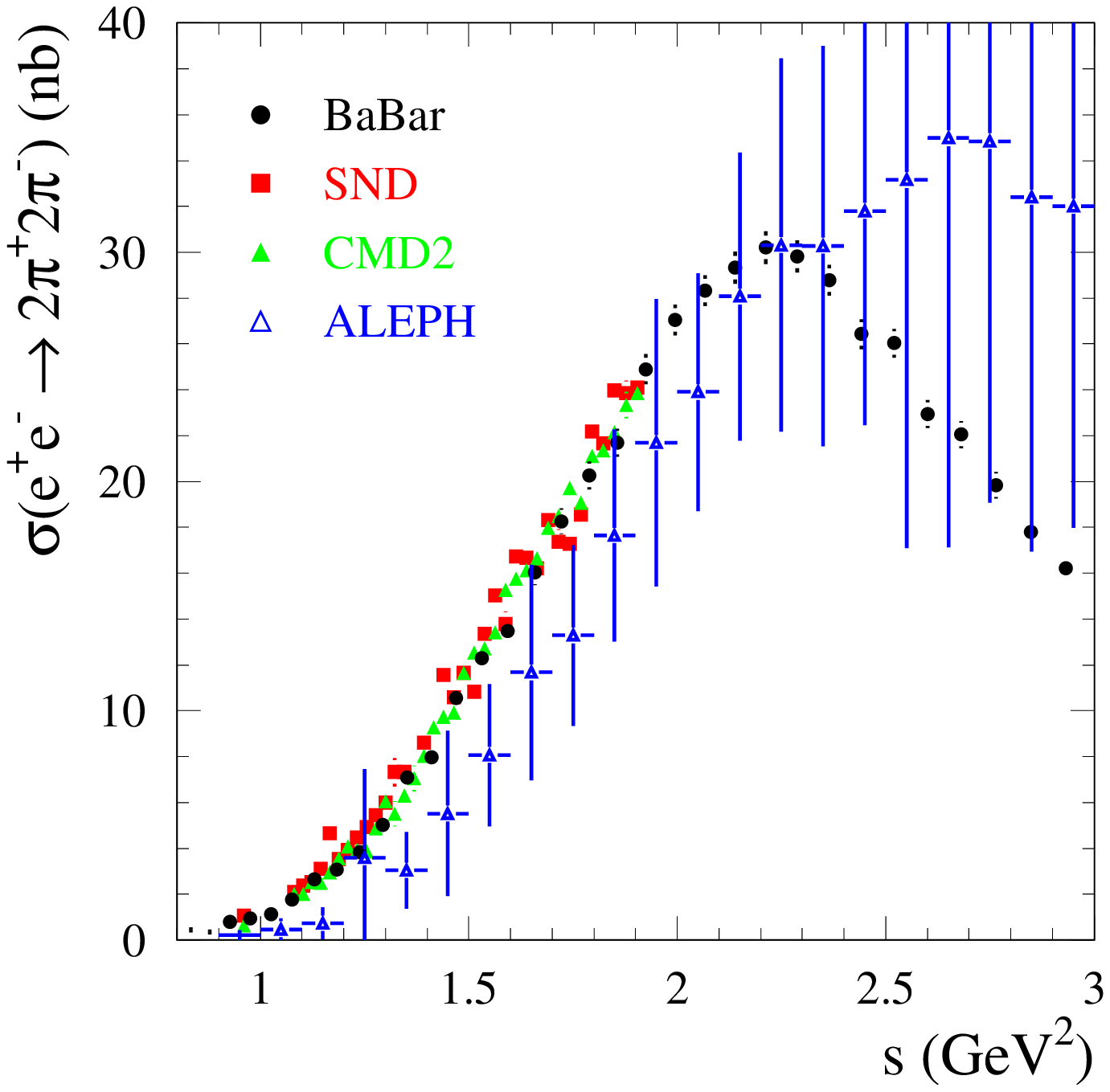}
\hfill
\includegraphics[width=.4\linewidth]{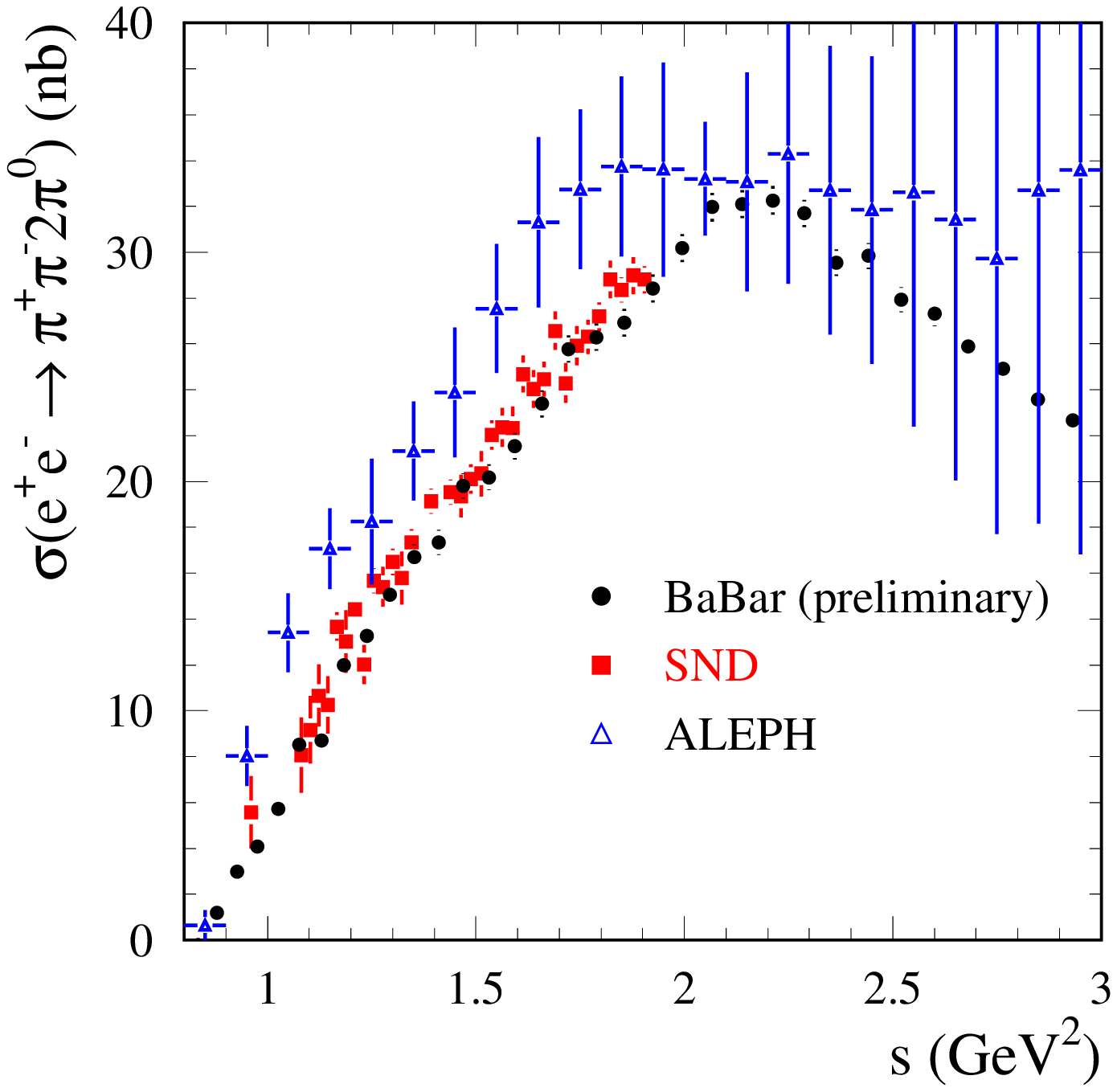}
\caption{Comparison of $e^+e^- \to 2\pi^+ 2\pi^-$ (left) and 
$e^+e^- \to \pi^+\pi^- 2\pi^0$ cross section measured in $e^+e^-$
collisions with the cross sections calculated from the spectral functions
of $\tau\to 4\pi \nu_\tau$ decays.
\label{tau4pi}}
\end{figure*}

The hypothesis of the vector current conservation (CVC) connects the
isovector $e^+e^-$ cross section with the vector spectral function
in $\tau$ decay:
$$\sigma^{I=1}_{e^+e^-\to X^0}(s)=\frac{4\pi\alpha^2}{s}v_{1,X^-}(s).$$
This relation holds in the limit of exact isospin invariance.
Therefore the $\tau$ spectral function must be corrected for
isospin-breaking effects before comparison with the corresponding $e^+e^-$
cross section. Such corrections were considered in Ref.~\cite{tau1}.
The pion form factor and $e^+e^-\to 4\pi$ cross sections calculated from
the isospin-breaking corrected $\tau$ spectral functions are compared with 
the direct $e^+e^-$ measurements in Figs.~\ref{tau2pi},\ref{tau4pi}.
Fig.~\ref{tau2pi} was taken from Ref.~\cite{g-2-th1}. The $\tau$ $2\pi$
spectral function is obtained by averaging the ALEPH, CLEO, and OPAL
data. The $\tau$ $4\pi$ spectral functions were taken from  Ref.~\cite{aleph}.
For all three reactions a significant discrepancy between $e^+e^-$ and 
$\tau$ data is observed. These disagreements are difficult to explain by
unaccounted isospin breaking effects. New precise $\tau$ data are needed
to confirm the observed discrepancies.

\section{Summary}
The significant progress has been reached in the measurements of the
low energy hadronic $e^+e^-$ cross sections in last few years,
leading to improvement of the accuracy of the total hadronic $e^+e^-$ cross
section. In particular, using the recent CMD-2, SND, KLOE, and BABAR data
reduces the error of the $a_\mu^{\rm had}$ from about $7\times 10^{-10}$
in 2003 to about $4\times 10^{-10}$ in 2007. This error is now comparable
with the error of the light-by-light scattering contribution, also $4\times
10^{-10}$~\cite{LBL}. 
  
There are, however, the discrepancy between the energy dependencies of
the pion form factors measured directly in $e^+e^-$ collisions at VEPP-2M and
via ISR at KLOE. The new ISR data on $e^+e^-\to\pi^+\pi^-$ are expected 
from the KLOE and BABAR. Both experiments plan to measure $\pi\pi/\mu\mu$
ratio. This will reduce the systematic error, in particular, due
to radiative effects.

The new precise data on multihadron cross sections for the energy region
from 1 to 3 GeV obtained at VEPP-2M and in BABAR experiment allows to 
improve our knowledge of the properties of the $\rho$, $\omega$, and $\phi$
excitations. All major isoscalar channels ($3\pi$, $\omega\pi\pi$, $\omega\eta$,
$KK^\ast$, $\phi\eta$) have been measured. Some theoretical input is needed to
perform a global fit to the measured cross sections and extract 
a complete set of parameters of the isoscalar resonances.
The many isovector 
cross sections ($2\pi^+2\pi^-$, $K^+K^-\pi\pi$, $\eta\rho$, $KK^\ast$) are 
also measured. The final BABAR result on $e^+e^-\to \pi^+\pi^- 2\pi^0$
is expected soon. 

The discrepancy between the $e^+e^-$ and isospin-breaking corrected data 
$\tau$ data is observed for $2\pi$ and $4\pi$ final states. 
This discrepancy is difficult to explain by
unaccounted isospin breaking effects. New data on $\tau\to 2\pi\nu_\tau$ and
$\tau\to 4\pi\nu_\tau$ decays from $B$ factories, and ISR data on $2\pi$ channel
from KLOE and BABAR should provide a test the existing $e^+e^-$ and $\tau$
measurements.

\end{document}